\documentclass[10pt,twocolumn]{article} 
\usepackage[pdftex,bookmarksnumbered=true]{hyperref}
\usepackage{latex8}
\usepackage{times}
\usepackage{helvet}

\usepackage[pdftex]{graphicx}

\usepackage[ruled, vlined]{algorithm2e}
\usepackage{verbatim}
\usepackage{amsmath, amsthm, amssymb}
\usepackage{pgf}

\newtheorem{definition}{Definition}

\newcommand{\adds}{\stackrel{\text{\tiny{+}}}{\gets}}
\newcommand{\zt}{\tilde{z}}
\usepackage{float}
\usepackage{amsmath,amsfonts}
\usepackage{amssymb}
\usepackage{amstext}
\usepackage[american]{babel}

\usepackage{xspace}
\usepackage{graphicx}
\usepackage{epsfig}
\usepackage{verbatim}
 \floatstyle{ruled}
 \newfloat{algorithm}{H}{alg}
 \floatname{algorithm}{Algorithm}
 \newfloat{figure}{H}{fig}
 \floatname{figure}{Figure}
 \newfloat{table}{H}{tab}
 \floatname{table}{Table}

\usepackage{ifpdf}
\ifpdf
\usepackage[pdftex]{graphicx}
\DeclareGraphicsExtensions{.pdf}
\else
\usepackage{graphicx}
\DeclareGraphicsExtensions{.eps}
\fi
\title{Managing Varying Worst Case Execution Times on DVS Platforms}

\author{Vandy {\scshape Berten}, Chi-Ju {\scshape Chang}, Tei-Wei {\scshape Kuo}\\
National Taiwan University\\
Computer Science and Information Engineering dept.\\
\{vberten, ktw\}@csie.ntu.edu.tw, james299kimo@gmail.com 
}

\begin{document}
\maketitle

\begin{abstract}
Energy efficient real-time task scheduling attracted a lot of attention in the past decade. Most of the time, deterministic execution lengths for tasks were considered, but this model fits less and less with the reality, especially with the increasing number of multimedia applications. It's why a lot of research is starting to consider stochastic models, where execution times are only known stochastically. However, authors consider that they have a pretty much precise knowledge about the properties of the system, especially regarding to the worst case execution time (or worst case execution cycles, WCEC). 

In this work, we try to relax this hypothesis, and assume that the WCEC can vary. We propose miscellaneous methods to react to such a situation, and give many simulation results attesting that with a small effort, we can provide very good results, allowing to keep a low deadline miss rate as well as an energy consumption similar to clairvoyant algorithms.
\end{abstract}

\section{Introduction}

\subsection{Motivations}
In the past decade, energy efficient systems have been actively explored. With the tremendous increase of the number of mobile devices, research in this field is still only at its beginning. Moreover, as many of those devices run now multimedia applications, real-time stochastic low-power systems will have a major role to play in the next few years.
One of the characteristic of stochastic systems, is that their parameters are likely to change in the time. For instance, a device decoding a movie will need more time to process a sequence in a movie with a lot of color and movement that a dark and quite sequence. What authors usually propose in the literature is to observe continuously the tasks execution time, in order to update their knowledge about the distribution. And once this distribution seems to be too far away from the one which was used by the scheduler, the scheduler is updated. But if the worst case execution time seems to increase, the system cannot afford to wait for collecting enough information in order to update the scheduler: some actions need to be taken as soon as possible, in order to avoid to miss deadline. This is what we want to do in this work: propose an efficient strategy to react to WCEC variation.

\subsection{Related work}
Energy-efficient real-time task scheduling attracted a lot of attention over the past decade. Low-power real-time systems with stochastic or unknown duration have been studied for several years. The problem has first been considered in systems with only one task, or systems in which each task gets a fixed amount of time. Gruian \cite{Gruian01,Gruian01b} or Lorch and Smith \cite{pace01,pace04} both shown that when intra-task frequency change is available, the more efficient way to save energy is to increase progressively the speed. 

Solutions using a discrete set of frequencies and taking speed change overhead into account have also been proposed \cite{Xu04,Xu07b}. For inter-task frequency changes, some work has been already undertaken. In \cite{Mosse00}, authors consider a similar model to the one we consider here, even if this model is presented differently. 
The authors present several dynamic power management techniques, with different aggressiveness level.
In \cite{Aydin01}, authors attempt to allow the manager to tune this aggressiveness level, while in \cite{Xu07b}, they propose to adapt automatically this aggressiveness using the distribution of the number of cycles for each task. 
The same authors have also proposed a strategy taking the number of available speeds into account from the beginning, instead of patching algorithms developed for continuous speed processors \cite{Xu07}. 
In~\cite{RTCSA08}, we generalize and uniformize the model presented in several of the previous papers, and propose two contributions: first, we gave a general sufficient and necessary condition of schedulability for this model, and second, we presented a new approach to adapt a continuous-speed-based method to a discrete-speed system.
Some multiprocessor extensions have been considered in \cite{Chen07}.

\subsection{Model}

The model considered in this document is the same as the one presented in \cite{RTCSA08} by the same authors. We resume it here for the sake of completeness, but the interested reader is invited to have a look at the given reference.
\begin{itemize}
   \item We have $N$ \emph{tasks} $\{T_i, i\in [1,\dots,N]\}$ which run on a DVS CPU. They all share the same deadline and period $D$ (also known as the \emph{frame}), do not have offset (periods are then synchronised) and are executed in the order $T_1$, $T_2$, \dots, $T_N$. The maximum execution number of cycles of $T_i$ is $w_i$;
  \item The \emph{CPU} can run at $M$ frequencies $f_1<\dots<f_M$; 
  \item We have $N$ \emph{scheduling functions} $S_i(t)$ for $i \in [1,\dots, N]$ and $t \in [0, D]$. This function means that if $T_i$ starts its execution at time $t$, it will run until its end --- unless the task is suspended before --- at frequency $S_i(t)$, where $S_i(t)\in \{f_1, f_2, ..., f_M\}$. $S_i(t)$ is then a step function (piece-wise constant function), with only $M$ possible values. 
\end{itemize}

\begin{figure}[!ht]
\begin{center}

\scalebox{0.68}{
\ifpdf
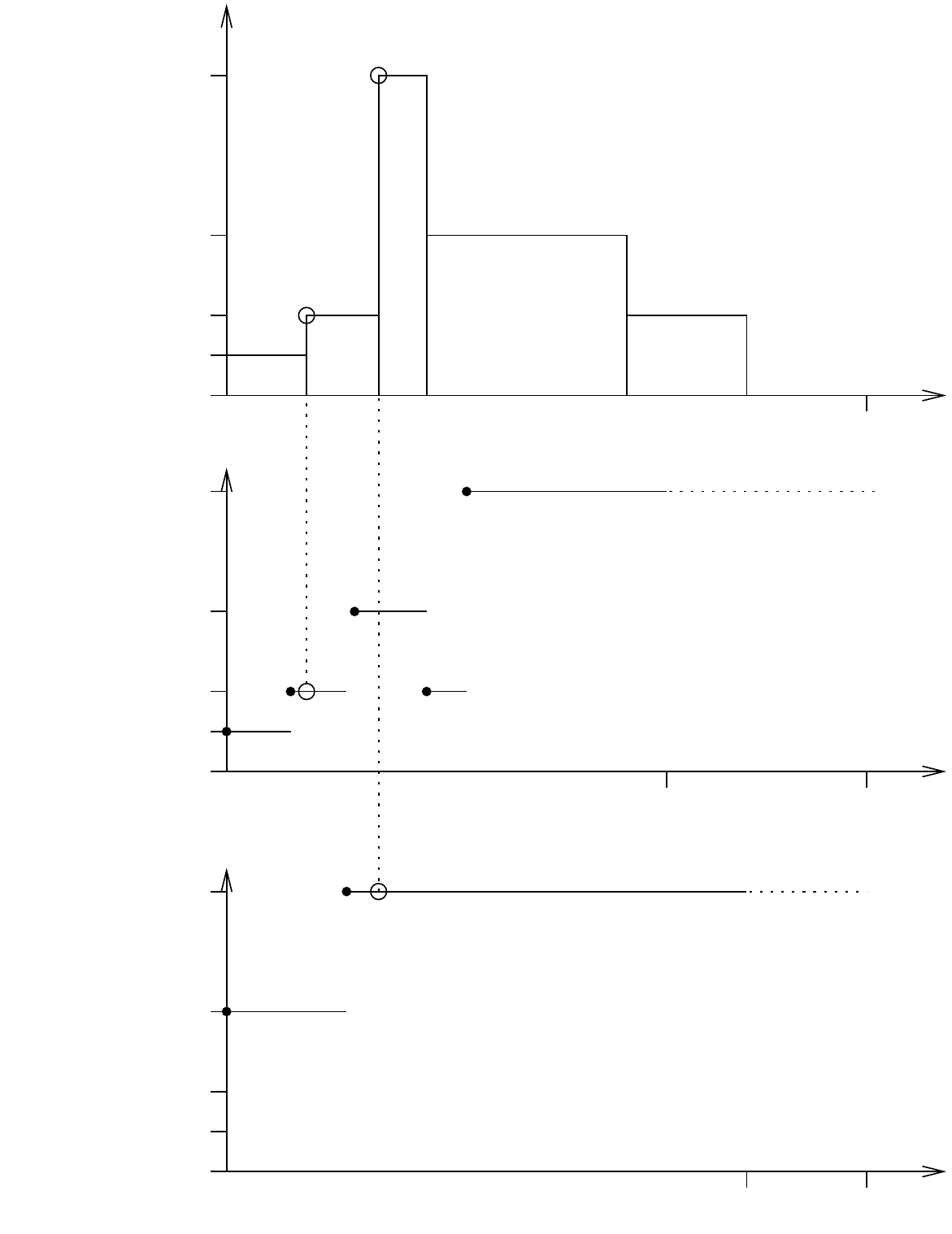
\else
\input{Sit.pstex_t}
\fi
}
\caption{\label{Sit} Example of scheduling with function $S_i(t)$. We have 5 tasks $T_1, \dots, T_5$, running every $D$. $T_1$ is run at frequency $f_1 = S_1(t_1)$, $T_2$ at $f_2 = S_2(t_2)$, $T_3$ at $f_4 = S_3(t_3)$, etc}
\end{center}
\end{figure}

This model generalizes several scheduling strategies proposed in the literature, such as  \cite{Xu07, Xu07b}.
Figure~\ref{Sit} shows an example of such scheduling function set.

A scheduling function can be modelized by a set of points (black dots on Figure~\ref{Sit}), representing the beginning of the step. $\mid S_i \mid$ is the number of steps of $S_i$. $S_i[k], k\in\{1, \dots, \mid S_i \mid\}$ is one point, with $S_i[k].t$ being its time component, and $S_i[k].f$ the frequency. $S_i$ has then the same value $S_i[k].f$ in the interval $\Big[S_i[k].t, S_i[k+1].t\Big[$ (with $S_i[\mid{}S_i\mid+1].t = \infty$), and we have 

$$
S_i(t) = S_i[k].f 
$$
where $k = \max \Big\{j\in \{1, \dots, \mid S_i \mid \} \Big|  S_i[j].t \le t\Big\}$

Notice that finding $k$ can be done in $O(\log \mid S_i \mid)$ (by binary search), and, except in the case of very particular models, $\mid S_i \mid \le M$.

Energy and time overhead for frequency changes can easily be taken into account in the model.

\subsubsection{Schedulability Condition}
The scheduling functions $S_i(t)$ can be pretty general, but have to respect some constraints in order to ensure the system schedulability and avoid deadline misses. If tasks never need more cycles than their worst case execution cycles (WCEC), or, in other word, if the knowledge about WCEC is correct, we show in \cite{RTCSA08} that the following (necessary and sufficient) condition ensures that every job in a frame can finish before the end of this frame (if the system is expedient and $T_1$ starts at the beginning of the frame): 

\begin{equation}\label{eq:sched}
S_i(t) \ge \dfrac{w_i}{z_{i+1}-t} ~\forall i \in [1, \dots, N], t \in [0, z_i[,
\end{equation}
 $$\text{where } z_i = D-\frac{1}{f_M}\sum_{k=i}^N w_k,$$

We don't take frequency change overheads into account, but we have shown in \cite{RTCSA08} that thoses penalties are easy to integrate.

\subsubsection{Danger Zone}
In \cite{RTCSA08}, we define the concept of \emph{Danger Zone}. The danger zone of a task $T_i$ starts at $z_i$, where $z_i$ is such that if this task is not started immediately, we cannot ensure that this task and every subsequent task can all be finished by the deadline (assuming WCEC are correctly known). In other words, if a task starts in its danger zone, and this task and all the subsequent ones use their WCEC, even at the highest frequency, some tasks will miss the (common) deadline.
The danger zone of $T_i$ is the range $]z_i, D]$, where

\begin{equation} \label{eq:DZ}
z_i = D-\frac{1}{f_M} \sum_{k=i}^N w_k.
\end{equation}

\subsubsection{Notation}
In this paper, we will use the following notation:

$$\lceil x \rceil_A = \left\{ 
\begin{array}{ll}
\min_{y\in A} \{y \big| y\ge x\} & \text{if } x \le \max \{A\} \\
\max_{y\in A} \{y\} & \text{otherwise}
\end{array}
\right.
$$

This notation generalizes the classical notation for the ceiling: $\lceil x \rceil = \lceil x \rceil_\mathbb{Z}$. For the sake of readability, $\lceil x \rceil_\mathcal{F}$ will denote $\lceil x \rceil_{\{f_1, \dots, f_M\}}$, or, in other words, the first frequency higher or equal to $x$, or $f_M$ if $x$ is higher than $f_M$.

Table~\ref{tab:symb} resumes symbols used in this document.

\begin{table}
\begin{center}
\begin{tabular}{l|l}
$c_i$		& Effective number of cycles of $T_i$\\ \hline
$D$		& Deadline \\ \hline
$f_i$		& Available frequencies\\ \hline
$M$		& Number of available frequencies\\ \hline
$N$		& Number of tasks \\ \hline
$\kappa_i(\varepsilon)$ & $\mathop{argmin}_\mathcal{K} \{ \mathbb{P}[c_i<\mathcal{K}] \ge 1-\varepsilon\}$ \\ \hline
$S_i(t)$	& Frequency if $T_i$ starts at $t$\\ \hline
$T_i$		& Task number $i$ \\ \hline
$w_i$		& Worst case execution cycles (WCEC) of $T_i$ \\ \hline
$z_i$		& Beginning of $T_i$ danger zone \\ 
\end{tabular}
\end{center}
\caption{Symbols used in this document.\label{tab:symb}}
\end{table}


\subsection{Varying WCEC}

Let us assume that we have a set of functions $S_i(\cdot)$, supposed the be as efficient as possible, thanks to any algorithm such as \cite{RTCSA08} or \cite{Xu07}. Those functions have been computed knowing the execution length distribution, as well as the worst case execution time $w_i$. 

It is realistic to consider that the distribution can vary according to the time. As long as $w_i$ does not change, there is no problem: from time to time, when the current distribution is too far away from the distribution used for the computation of $S_i(\cdot)$, those functions are rebuild. The computation of those functions can for instance be done during the slack time of any frame, or on another CPU. Meanwhile, the ``old'' functions can still be used without risking deadline violation. In the worst case, deprecated functions can make the system consuming more energy than an up-to-date set of functions.

But the WCEC can also vary. If a $w_i$ decreases, again, this does not have any consequence on the schedulability. If this $w_i$ increases, this means that suddenly, an instance of $T_i$ runs over the maximal time it was supposed to run. Here is the aim of this work: study miscellaneous possible reactions to this situation, allowing to reduce as much as possible deadline misses or task killings (which in some cases are intrinsically unavoidably), before the system can rebuild its functions.

The strategies we propose here are composed of two phases:
\begin{itemize}
 \item First, we have to decide what to do with tasks overpassing their $w_i$ in the current frame;
 \item Then, we have to adapt the function $S_i(\cdot)$ in order to guarantee the schedulability of the next frame taking into account the new $w_i$ (if other tasks won't be longer than expected. Otherwise, again, the schedulability is not guaranteed). Of course, we do not completely rebuild $S_i(\cdot)$ which would take too much time. Instead, we change them slightly, improving schedulability, but possibly reducing power consumption efficiency.
\end{itemize}

In this work, we will assume that we cannot tolerate that at time $D$, some tasks of the current frame are still running. This means that we prefere to kill some tasks instead of missing deadlines. We need then a mechanism allowing to kill a task if still running by some given time that can be provided at the beginning of this task execution.

\section{No preemption mechanism}
We first assume that tasks cannot be preempted: tasks can then be killed, but if at the end of the frame, it happens that there is some time left, the killed task cannot be resumed. We consider that a situation becomes problematic once a task enters (or does not end before) the danger zone of the next task. 
We have mainly three solutions:
\begin{itemize}
 \item We do not want any subsequent task to enter the next danger zone. We kill then the running task just before the danger zone of the next task;
 \item We let the task running without any control. At time $D$, we kill any task that would be still running;
 \item An hybrid solution.
\end{itemize}
\subsection{Kill at danger zone}
In this case, we consider that a task running longer than expected is ``guilty'', and its then killed as soon as it could make any task to be not able to meet its deadline. When task $T_i$ is started, a timeout is set (and disabled when the task finishes) on $z_{i+1}$ (see Equation~\ref{eq:DZ}), the start of $T_{i+1}$ danger zone (or $D$ if $i=N$).

The main disadvantage is that the danger zone is usually very pessimistic. There is then a high probability (depending upon the computation cycles distribution) that the system kills $T_i$, but some time is still available after $T_N$ (except, of course, if $i=N$).

Another disadvantage of this technique is that, as the next task $T_{i+1}$ starts at the limit of its danger zone, it will be killed as soon as it uses more cycles than $w_{i+1}$. So killing a task ``cancels'' the laxity of the next one.

\subsection{Kill at \texorpdfstring{$D$}{D}}
Here, we accept a task to enter in the next danger zone. Of course, if a task starts during its danger zone, it should run as fast as possible, then,
$$
S_i(t) = f_M ~\forall t > z_i 
$$

At the beginning of a frame, at timeout is set on $D$, and if a task is still running at this time, this task is killed, and any subsequent task is just dropped. The main advantage of this technique is that it allows to use the totality of the available time. Then the ``time overflow'' has to be huge before a task is killed. 

However, the last task of a frame is always expected to run at a speed which allows it to use the totality of the remaining time if it uses its WCEC. Indeed, if the frequency cannot be changed in the middle of a task, this is usually the speed giving the minimum energy consumption. 
Then, the laxity for the tasks at the beginning of a frame is larger than the laxity at the end, which could be considered as unfair.

Another disadvantage is that if a task is killed or not run, the last one does obviously not reach its end, even if it does not require more than $w_N$. Then, the larger $i$, the larger the probability to be killed or not run. Again, this is unfair.

One possibility is to put at the end of the frame tasks that are less important, or even optional.

\subsection{Hybrid solutions}
An hybrid solution is to give more space to the last tasks but still starting some tasks in their danger zone expecting them to run less than their WCEC. We allow any task to enter the danger zone of the next task, but the large $i$, the less intrusion is allowed. In a strict solution, $T_i$ would be killed at $z_{i+1}$. We want to define a set of points $\{\zt_i\}$ such as $T_i$ is killed at $\zt_{i+1}$, with the following properties:
\begin{itemize}
   \item $\zt_i \ge z_i$: we are less strict than just killing at the next danger zone;
   \item $\zt_i < \zt_{i+1}$: when a task is killed, the next one needs to have some time left;
   \item $\zt_{N+1} = D$: $T_N$ is killed at $D$ if still running (because of firm deadlines);
\end{itemize}

We propose two methods allowing this: the first one allows a task $T_i$ to use a fixed ratio of the time interval $[z_{i+1}, D]$, the second one takes the job length distribution into account.

The first one defines then 
$$\zt_i = z_i + (D-z_i) \delta_i$$
for some $0\le \delta_i \le 1$. We can see easily that $\delta=1$ corresponds to ``Kill at D'', and $\delta=0$ at ``kill at Danger Zone''.

\subsubsection{Percentile approach}\label{sect:vareps}

An alternative hybrid solution is to consider that any of the subsequent tasks will not need more than its $1-\varepsilon$ percentile. If $\kappa_i(\varepsilon)$ $(\in [0, w_i]) = \mathop{argmin}_\mathcal{K} \{ \mathbb{P}[c_i<\mathcal{K}] \ge 1-\varepsilon\}$, ($\kappa_i(\varepsilon)$ is then the length which is not overpassed by $(1-\varepsilon)\times 100\%$ of jobs) where $c_i$ is the effective number of used cycles, then we only accept $T_i$ to run up to 

$$
\zt_{i+1} = D - \frac{1}{f_M} \left(\sum_{j=i+1}^N \kappa_j \right).
$$

In the following, for clarity, $\kappa_i$ stands for $\kappa_i(\varepsilon)$. For instance, we set $\varepsilon$ to $5\%$, and, using the $T_{k}$ number of cycles distribution, we compute $\kappa_k$ such as in $95\%$ of cases, $T_k$ uses less than $\kappa_k$ cycles. Then we behave as if $\kappa_k$ was the WCEC of $T_k$ for each $k>i$. 


The tolerance $\varepsilon$ does not have to be the same for every task: we could set $\varepsilon=0$ for some important tasks, and $\varepsilon_i<\varepsilon_j$ if $i$ has an higher priority than $j$.

In order to improve the fairness between $T_N$ and other tasks, we can also consider a decreasing sequence of $\varepsilon$.

However, this technique could be considered as unfair: first tasks have much more laxity than last tasks. A way of working around this would be to consider the $1-\varepsilon$ percentile only for some fixed number of tasks, say $K$, and using the following value:
$$
\zt_{i+1} = D - \frac{1}{f_M} \left(\sum_{j=i+1}^{\min\{i+K, N\}} \kappa_j + \sum_{j=i+K+1}^N w_j \right).
$$

\section{Available preemption mechanism}

If tasks can be preempted, we can then suspend a task as soon as it enters the next danger zone (or later, at $\zt_{i+1}$), and resume it:
\begin{itemize}
   \item Either after the end or the preemption of $T_N$. Of course, there is no guaranty that $T_N$ finishes before $D$, but we can assume that in general, it is highly likely;
   \item Or as soon as some slack is available, which means that a tasks $T_i$ ends before $\zt_{i+1}$. 
\end{itemize}

There are in this case several questions:
\begin{itemize}
   \item At which speed the resumed task should be run?
   \item What if several tasks have been suspended?
\end{itemize}

\subsection{Frequency for resumed task (resume at the end)}
The safest method would be to run a resumed task at speed $f_M$, but it will consume a lot of energy.
In some cases, we might have more information. First, we could know the global worst case execution cycle number $W_i$, such as $w_i\le W_i$. In this case, we can choose the lowest frequency allowing to run the remaining part of $T_i$ assuming it could use $W_i$ in total.
Then, if $T_i$ is the only task to run at time $t$, if $c_i$ is the number of cycles already consumed, we can choose


$$
f = \left\lceil \dfrac{W_i-c_i}{D-t} \right\rceil_\mathcal{F}.
$$

If $f_M < \frac{W_i-c_i}{D-t}$ (then $f=f_M$), this means that even at $f_M$, we cannot guarantee to finish the task before $D$. But this does not mean that we are going to miss the deadline, because the task can of course require less cycles than $W_i$. In such a situation, if not missing deadline is more important than energy consumption, $f_M$ is the best frequency.


In some situations, we could for instance assume that the WCEC does not vary too abruptly. In other terms, if $w_i$ is the current WCEC, a task will never need more that $w_i \cdot (1+\alpha)$, for some $\alpha>0$, or at least we accept to kill a task if this condition is not met. In that case, a similar frequency selection can be used:

$$f = \left\lceil \dfrac{w_i \cdot (1+\alpha)-c_i}{D-t} \right\rceil_\mathcal{F}$$

\subsection{Frequency for resumed task (resume at the first slack)}
Again, the most conservative way is to resume tasks at their maximal speed. This conservative method is of course the safest, but consumes more energy. The disadvantage of resuming tasks as soon as we have some available slack is that we don't really know the available total slack. We only know the time we have before the next tasks start, but some slack might be available later. So several heuristics might be considered:
\begin{itemize}
   \item Resume tasks at the current speed;
   \item Resume tasks at $f_M$
   \item According to the ``emergency'' (current available slack, number of tasks to be run, probability to find some slack later, \dots), choose a frequency.
\end{itemize}

\subsection{Frequency for other tasks}

If we choose to resume tasks after $T_N$, and if a task has been suspended, a first method would be to just ``forget'' about this task before the end of $T_N$, which means choose the frequency for subsequence tasks without taking this suspended task into account. But in order to reduce the power consumption, the system should select the smallest frequency allowing $T_N$ to finish just in time if is uses its WCEC. If the execution number of cycles variance is rather small, the slack time could be pretty much tiny. Especially because if the user has the choice, it is in general better to put tasks with smaller variance at the end, because this allows to finish the last task very close to the deadline, and then reduce the idle time, and have a better repartition.

A solution would be to increase the speed of tasks as soon as some tasks are waiting in the ``suspended tasks queue''. The most conservative way is that we stop using $S$ function when tasks are suspended, and always use $f_M$. A more optimistic way would be to increase the speed that $S$ would have chosen, for instance the next available frequency, or + the number of suspended tasks.

\subsection{Several tasks}

If several tasks have been interrupted and should be resumed, the situation is slightly more complex, and some strategy needs to be defined. Let $R$ be the set of tasks to resume.

First, if we do not have any information about the maximal number of remaining cycles, the safest way is to run tasks at $f_M$. 
We also have to choose in what order tasks from $R$ are resumed. The simplest way is to resume them in the order given by indices. But this give some fairness problems: tasks with a higher index have a higher likelyhood to be killed. 
This could fit the user requirements if tasks are sorted according to their priority, but not in case of uniform priorities.

In order to improve the fairness, a simple method consists in resuming tasks randomly, possibly with some criteria according to the user needs.

If we have for any task in $R$ the knowledge of its global maximum $W_i$ (with possibly $W_i=w_i \cdot (1+\alpha)$), we can then use the frequency

$$f = 
	\left\lceil 	
		\dfrac{\sum_{i \in R} (W_i-c_i)}{D-t}  
	\right\rceil_\mathcal{F}.$$

Of course, for better energy consumption, this frequency has to be recomputed before resuming any task of $R$, because some previous resumed tasks could have required less time than their expected maximum.

\subsubsection{Multiple rounds}
In the previous section, we only consider situations where resumed tasks are not suspended later. But we can also consider that tasks can be interrupted or resumed several times. We then have to wonder if we would better to finish the maximum number of tasks, or to progress in every task in parallel. It might be interesting to consider two families of tasks:
\begin{itemize}
   \item If a not finished tasks does not give any benefit, the time spent in running such tasks is just wasted. In this case, it would be probably better to run a task up to its completion;
   \item If a not finished can give some feedback (for instance with lost of quality and/or precision, as it is the case in the Imprecise Computation Model~\cite{Leung07}), then we could try to improve the fairness instead of the number of finished tasks.
\end{itemize}

In the first case, we want to maximize to number of finished tasks. If we do not have any information about the remaining time, then any order might be equivalent if there are not priority between tasks. If we have some information about the remaining time, then we could heuristically resume tasks sorted by then smallest remaining time first.

In the second case, we could prefer to improve the fairness by running tasks in parallel. Of course, we are only interested by the fairness at time $D$, then we do not need ``real'' parallelism. For instance, we can allocate to each task in $R$ the same amount of time. Then, if $T_N$ ends at $t'$, every task gets $(D-t')/\parallel R\parallel$ (or any weighted repartition). If a task reaches its allocated amount of time, it is re-suspended. If some tasks needed less than their allocated time, a second round of resumes can be performed. This second round contains necessarily less tasks, as it is because some tasks end before their allocated time that we still have some left time. Notice that the worst number of preemptions occurs when at each round, one task finishes earlier than the time it receives. So if $r=\mid R\mid$, the number of preemptions is $(r-1)\times(r-2)\times\dots\times2 = (r-1)\times r/2$.

In order to avoid to have too much switching times, we could possibly decide (arbitrarily) to run only one task if the remaining time goes bellow some bound.

\subsection{Preemption and intra-task frequency changes}

If a preemption mechanism is available, we can use it in order to increase the frequency of a task. For tasks respecting their WCEC, we assume that scheduling functions do not give any information allowing to optimally change the frequency during execution. So if a task does not need more than its WCEC, we do not consider changing its speed in this work.

But if some task does not respect its expected delay, we can the interrupt it, and resume it immediately at a higher frequency. This allows to be safer, especially if we do not have information about the global WCEC. Several policies can be considered, the safest being to use $f_M$ as soon as a WCEC is not respected. But this strategy is certainly the most energy consuming.

We can also consider to take into account the laxity we have, that is, the remaining time before the next danger zone. We can for instance increase the frequency as the laxity diminishes.

\section{Adaptation of \texorpdfstring{$S_i(\cdot)$}{Si(.)}}

If $T_i$ needed $c_i >w_i$ cycles in the previous frame, then $c_i$ might be considered as the new WCEC (the case where $T_i$ has been killed after $c_i$ cycles is considered later). Of course, it is usually not realistic to rebuild the whole set of scheduling functions $S_i(\cdot)$ before the next frame. We choose then to adapt the scheduling function in order to guarantee the schedulability, where the WCECs are now $w_i^\text{new} = \max\{c_i, w_i\}$.

Let first consider that only one task $T_j$ overran its WCEC. 

The scheduling has now to guarantee that even if $T_j$ requires again $c_j$ cycles, it will end before $z_{j+1}$. We know that $S_j(\cdot)$ guarantees that if $T_j$ requires $w_j$ cycles, it will end before $z_{j+1}$. So we need to build a set of functions $S'_i(t)$ such as (see Equation~\ref{eq:sched}):

\begin{itemize}
 \item If $i < j$: 
\begin{equation}\label{eq:Sit1}
S'_i(t) \ge \frac{w_i}{D - t - \frac{1}{f_M} \left(c_j + \sum\limits_{\begin{subarray}{c}k=i+1\\ k\ne j\end{subarray} }^N w_k \right)}\end{equation}
 
 knowing that
$$ 
S_i(t) \ge \frac{w_i}{D - t - \frac{1}{f_M} \left(w_j + \sum\limits_{\begin{subarray}{c}k=i+1\\ k\ne j\end{subarray} }^N w_k \right)};
$$
\item If $i = j$:
\begin{equation}\label{eq:Sit2}
S'_j(t) \ge \frac{c_j}{z_{j+1} - t}\text{~~ knowing that ~~} S_j(t) \ge \frac{w_j}{z_{j+1} - t};
\end{equation}
\item If $i>j$:
\begin{equation}\label{eq:Sit3}
S'_i(t) \ge \frac{w_i}{z_{i+1} - t}\text{~~~ knowing that ~~~} S_i(t) \ge \frac{w_i}{z_{i+1}- t}.
\end{equation}
\end{itemize}

For $i>j$, we can simply take $S'_i(t) = S_i(t)$. The lower bound of $S_i(t)$ does not depend on tasks running before $T_i$.
We will now propose two adaptation methods for $i\le j$.

\subsection{Using Schedulability Condition}

A first adaptation consists in using the property that we want a function $S'_i(t)$ close to $S_i(t)$, but respecting the schedulability condition $S'_i(t) \ge \dfrac{c_j}{z_{i+1}-t}$. We propose then to simply use $S_i(t)$ as long as it respects the condition, and to use this condition when required. Or, more formally,

\begin{eqnarray*}
S'_i(t) &=& \left\lceil \max\left\{S_i(t), \frac{c_j}{z_{i+1}-t}\right\} \right\rceil_\mathcal{F} \\
&=& \max\left\{S_i(t), \left\lceil \frac{c_j}{z_{i+1}-t}\right\rceil_\mathcal{F} \right\} 
\end{eqnarray*}


Then, if a task $T_j$ took $c_j>w_j$ cycles in this frame, we make the following changes for the next frame:
\begin{itemize}
   \item $w_j \gets c_j$;
   \item Functions $S_i(\cdot)$ are changed in the following way:
   
$$
S'_i(t) \gets \left\{
\begin{array}{ll}
\max\left\{S_i(t), \left\lceil \dfrac{c_j}{z_{i+1}-t}\right\rceil_\mathcal{F} \right\}  & \text{if $i\le j$,} \\ \\
S_i(t) & \text{if $i>j$,}
\end{array}
\right.
$$
\end{itemize}

Adaptation of $S_i$ for $i \in [1, \dots, j]$ can be done using Algorithm~\ref{alg:adaptSSched}. 

\begin{algorithm}[ht]
\label{alg:adaptSSched}
\caption{Adaptation of $S_i$ (schedulability condition)}
\ForEach{$p \in \{2, \dots, \mid S_i\mid \}$}{
	\If{$S_i[p].f < \dfrac{c_j}{z_{i+1} - S_i[p].t}$}{
		$S_i[p].f \gets \left\lceil \dfrac{c_j}{z_{i+1} - S_i[p].t}\right\rceil_\mathcal{F}$ \;
	}
	\If{$S_i[p-1].f < \dfrac{c_j}{z_{i+1} - S_i[p].t}$}{
		\eIf{No frequency available between $S_i[p].f$ and $S_i[p-1].f$}{
			$S_i[p].t \gets z_{i+1} - \dfrac{c_j}{S_i[p-1].f}$ \;
		}{
			$S_i \adds \left(z_{i+1} - \dfrac{c_j}{S_i[p-1].f}, \left\lceil S_i[p].f^+ \right\rceil_\mathcal{F} \right)$ \;
			$p\gets p-1$ \;
		}
	}
}
\end{algorithm}

\subsection{Horizontal shift}

Another adaptation can be done using some properties of the function.
For $i<j$, we propose to define $S'_i(t) = S_i\left(t+\dfrac{c_j-w_j}{f_M}\right)$. 

This corresponds to a left shift of the amount of time required by the ``new cycles'', or ($c_j-w_j$).
Let us now check the schedulability condition of $S'_i$ for $i\le j$. 

We have 
\begin{eqnarray*}
S'_i(t) &=& S_i\left(t+\dfrac{c_j-w_j}{f_M}\right) \\
	&\ge& \frac{w_i}{D - t - \dfrac{1}{f_M} \left(c_j-w_j+w_j + \sum\limits_{\begin{subarray}{c}k=i+1\\ k\ne j\end{subarray} }^N w_k \right)}\\
	&=& \frac{w_i}{D - t - \dfrac{1}{f_M} \left(c_j+ \sum\limits_{\begin{subarray}{c}k=i+1\\ k\ne j\end{subarray} }^N w_k \right)}
\end{eqnarray*}
which is the condition \eqref{eq:Sit1}.

Combining the previous adaptation for $i=j$ and this one for $i<j$, $S_i$ can be adapted in the following way: 

$$
S'_i(t) \gets \left\{
\begin{array}{ll}
S_i\left(t+\dfrac{c_j-w_j}{f_M}\right) & \text{if $i<j$,}\\ \\ 
\max\left\{S_i(t), \left\lceil \dfrac{c_j}{z_{i+1}-t}\right\rceil_\mathcal{F} \right\}  & \text{if $i=j$,} \\ \\
S_i(t) & \text{if $i>j$,}
\end{array}
\right.
$$

This adaptation can be done in $O(j \times M+M \log M)$ (if $\mid S_i \mid \le M \forall i$), both numbers being usually very small, with Algorithm~\ref{alg:adaptS_Shift}. Notice that the logarithmic complexity is because of the ``ceiling'' (we need to find the smallest frequency higher than a value). But in practice, we don't have to consider all frequencies, because the new $S_j[k].f$ is obviously higher than the previous one. And if $c_j$ is not too different from $w_j$, the new frequency will simply be the next one, or the next to next one.

\begin{algorithm}[ht]
\label{alg:adaptS_Shift}
\caption{Adaptation of $S_i (i<j)$ (horizontal shift)}
\ForEach{$p \in \{1, \dots, \mid S_i\mid \}$}{
	$S_i[p].t \gets \max\left\{0, S_i[p].t - \dfrac{c_j-w_j}{f_M}\right\}$\;
}
\end{algorithm}

If several tasks overpass their WCEC in the last frame, we can just consider that this last frame is equivalent to as much frames as the number of tasks having overpassed their WCEC, and in each of those frames, only one task overpassed its WCEC.
We can then just apply the transformation given hereabove successively, once for each task.

Remark that if a task has been killed after using $c_i$, it may happen that the new WCEC is actually larger that $c_i$. So we may consider a value higher than $c_i$ as the new WCEC. However, in the next frame, there is a non null probability that $T_i$ still has some laxity after $c_i$ cycles in the next frame. This makes that, stochastically certainly, the system will converge to the new correct WCEC, possibly after having been killed again.

\subsection{Adapting Killing/Suspend Time}
Let $\zt_i$ be the time at which $T_{i-1}$ is killed (or suspended). According to the relationship between $\zt_i$ and $z_i$, $\zt_i$ must also be adapted when some WCEC changes.
Let $z'_i$ (resp. $\zt'_i$) be the danger zone (resp. kill/suspend time) after the change. If $j$ is the index of the task increasing its WCEC to $c_j$, we have

$$
z_i = D - \frac{1}{f_M} \sum_{k=i}^N w_i.
$$

and 

\begin{eqnarray*}
z'_i &=& D - \frac{1}{f_M} \sum_{\substack{k=i \\i\ne j}}^N w_i- \frac{1}{f_M} c_j \\
	&=& \left\{
	 \begin{array}{ll}
	 z_i - \dfrac{c_j-w_j}{f_M}	& \text{if }i\le j \\
	 z_i 				& \text{otherwise}   
	 \end{array} \right.
\end{eqnarray*}
Then, the new danger zone can be obtained by subtracting $\dfrac{c_j-w_j}{f_M}$ when $i\le j$.

If $\zt_i =  z_i + (D-z_i) \delta_i$ (which includes ``kill/suspend at danger zone'' and ``kill/suspend at $D$''), we have, if $i\le j$:

\begin{eqnarray*}
\zt'_i &=& z'_i + (D-z'_i) \delta_i \\
	&=& z_i - \frac{c_j-w_j}{f_M} + \left(D - \left(z_i - \frac{c_j-w_j}{f_M}\right)\right) \delta_i \\
	&=& z_i  + \left(D - z_i \right) \delta_i - \frac{c_j-w_j}{f_M} + \frac{c_j-w_j}{f_M}\delta_i \\
	&=& \zt_i - (1-\delta_i) \frac{c_j-w_j}{f_M}
\end{eqnarray*}

The new kill/suspend time is then obtained by subtracting $(1-\delta_i) \dfrac{c_j-w_j}{f_M}$ for $i \le j$.

The percentile approach is a little bit more difficult to solve, because this method is based on the task length distribution, and we assume that the known distribution is obsolete when we need to adapt $S$ functions, and therefore we should not use it anymore. We need then to adapt the kill/suspend time according to some heuristic.

We propose two transformations for $\kappa_j$ (we assume $c_j > w_j$):
\begin{enumerate}
   \item $\kappa'_j =  \kappa \dfrac{c_j}{w_j}$
   \item $\kappa'_j =  \kappa + (c_j - w_j)$
\end{enumerate}
The first adaptation assumes that the whole distribution is stretched from $[0, w_j]$ to $[0, c_j]$. The second adaptation assumes that the distribution is shifted upwards with a shift of $(c_j-w_j)$. We consider the generic general percentile approach:

$$
\zt_i = D - \frac{1}{f_M} \left(\sum_{k=i+1}^{\min\{i+K-1, N\}} \kappa_k + \sum_{k=i+K}^N w_k \right).
$$

The first adaptation gives:

$$
\zt'_i = \left\{\begin{array}{l}
            D - \frac{1}{f_M} \left(\sum\limits_{k=i}^{m_i^K} \kappa_k + \kappa_j(\frac{c_j}{w_j} -1)+ \sum\limits_{k=i+K}^N w_k \right) \\ \\
            \hfill \text{if }j\le m_i^K \\ \\
            D - \frac{1}{f_M} \left(\sum\limits_{k=i}^{m_i^K} \kappa_k + \sum\limits_{k=i+K}^N w_k + (c_j-w_j)\right)\\ \\
            \hfill \text{otherwise} 
         \end{array}
         \right.
$$
where $m_i^K = \min\{i+K-1, N\}$, or 

$$
\zt'_i = \left\{\begin{array}{ll}
            \zt_i - \dfrac{\kappa_j}{f_M} (\frac{c_j}{w_j} -1) & \text{if }j\le m_i^K \\ \\
            \zt_i - \dfrac{c_j-w_j}{f_M} & \text{otherwise} 
         \end{array}
         \right.
$$

If we consider the second adaptation, we can easily get:

$$
\zt'_i = \zt_i - \dfrac{c_j-w_j}{f_M} 
$$

\subsection{WCEC diminuing}

If the WCEC for some job has not been observed for a long time, we could consider to reduce it. Let $c_j$ ($<w_j$) be the ``new'' worst case execution cycle of $T_j$. While there is no schedulability necessity to adapt the scheduling, we could try to take this information into account in order to reduce the consumption. Again, we want to have simple changes before recomputing correctly scheduling functions.

Of course, we still need to ensure the schedulability (according to the new WCEC set). For instance, we could use an adaption very close to the one we did before:

$$
S'_i(t) \gets \left\{
\begin{array}{ll}
S_i\left(t-\dfrac{w_j-c_j}{f_M}\right) & \text{if $i<j$,}\\ \\ 
\left\lceil S_j(t) \dfrac{c_j}{w_j} \right\rceil_\mathcal{F} & \text{if $i=j$,} \\ \\
S_i(t) & \text{if $i>j$,}
\end{array}
\right.
$$

In the case $i<j$, we want to do a right shift. The schedulability of the case $i=j$ is very easy to show. The schedulability for other cases can be proved in a very similar way as before. Algorithm~\ref{alg:adaptS_Shift} can be used with a slight adaptation.

\section{Simulation Results}
   
\subsection{Scenario}

\begin{figure*}
   \begin{center}
   \newcommand{\lw}{0.24\linewidth}
      \includegraphics[width=\lw]{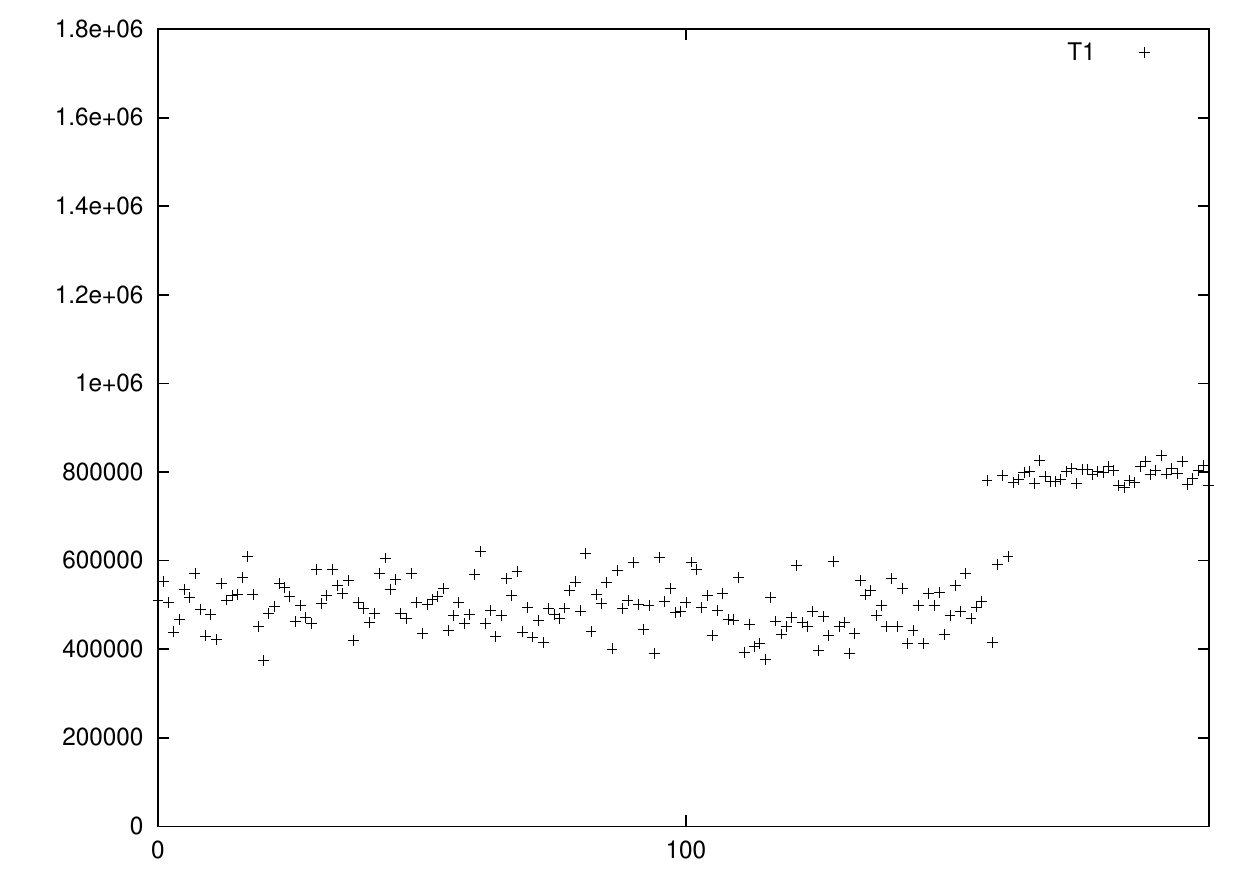}
      \includegraphics[width=\lw]{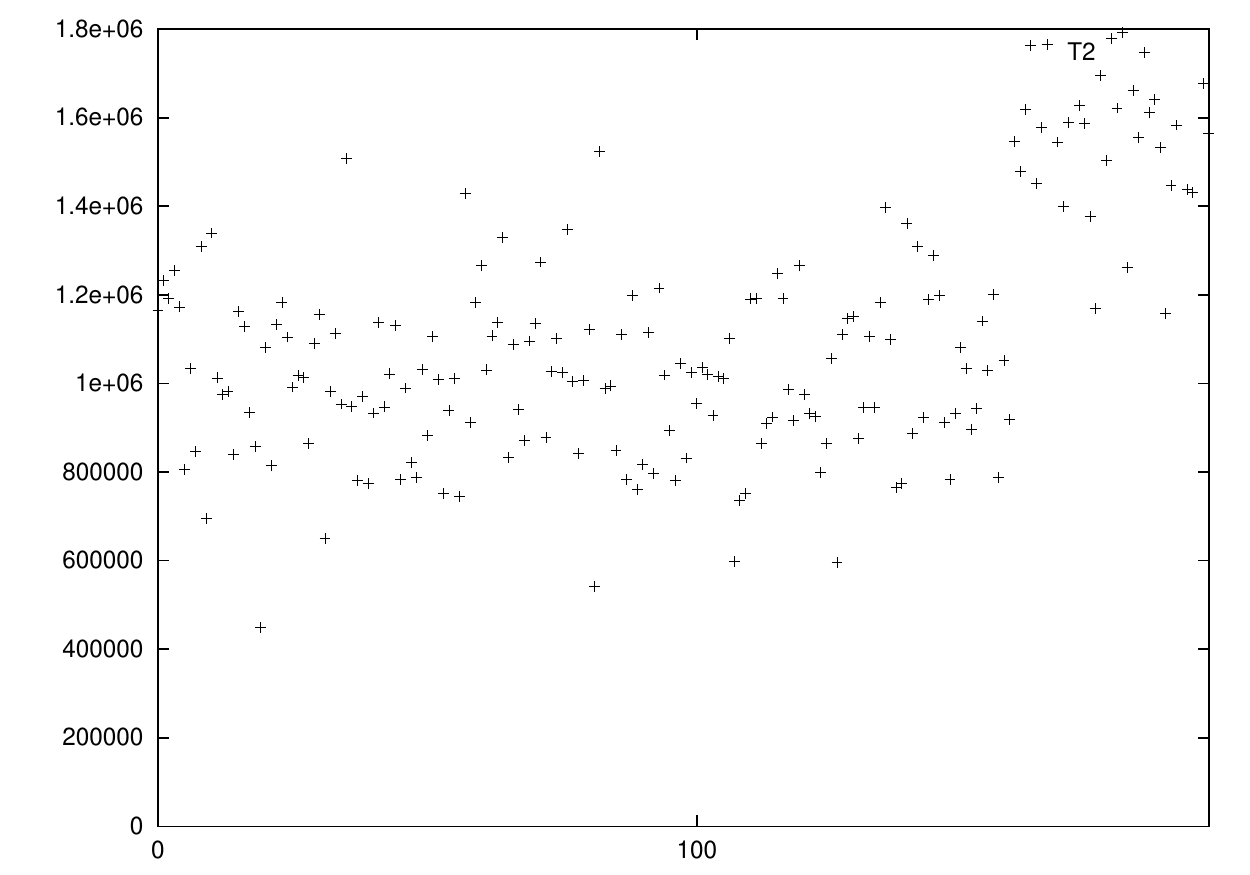}
      \includegraphics[width=\lw]{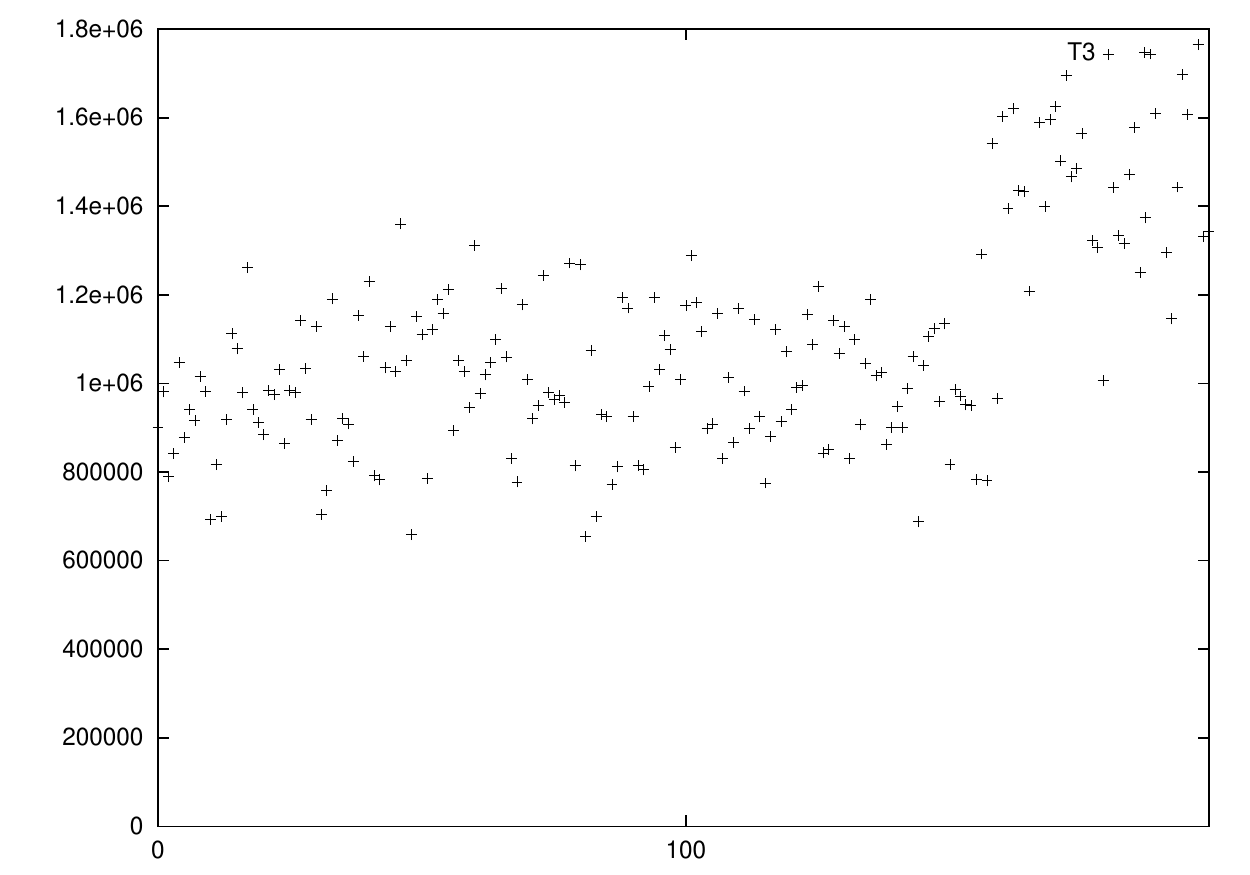}
      \includegraphics[width=\lw]{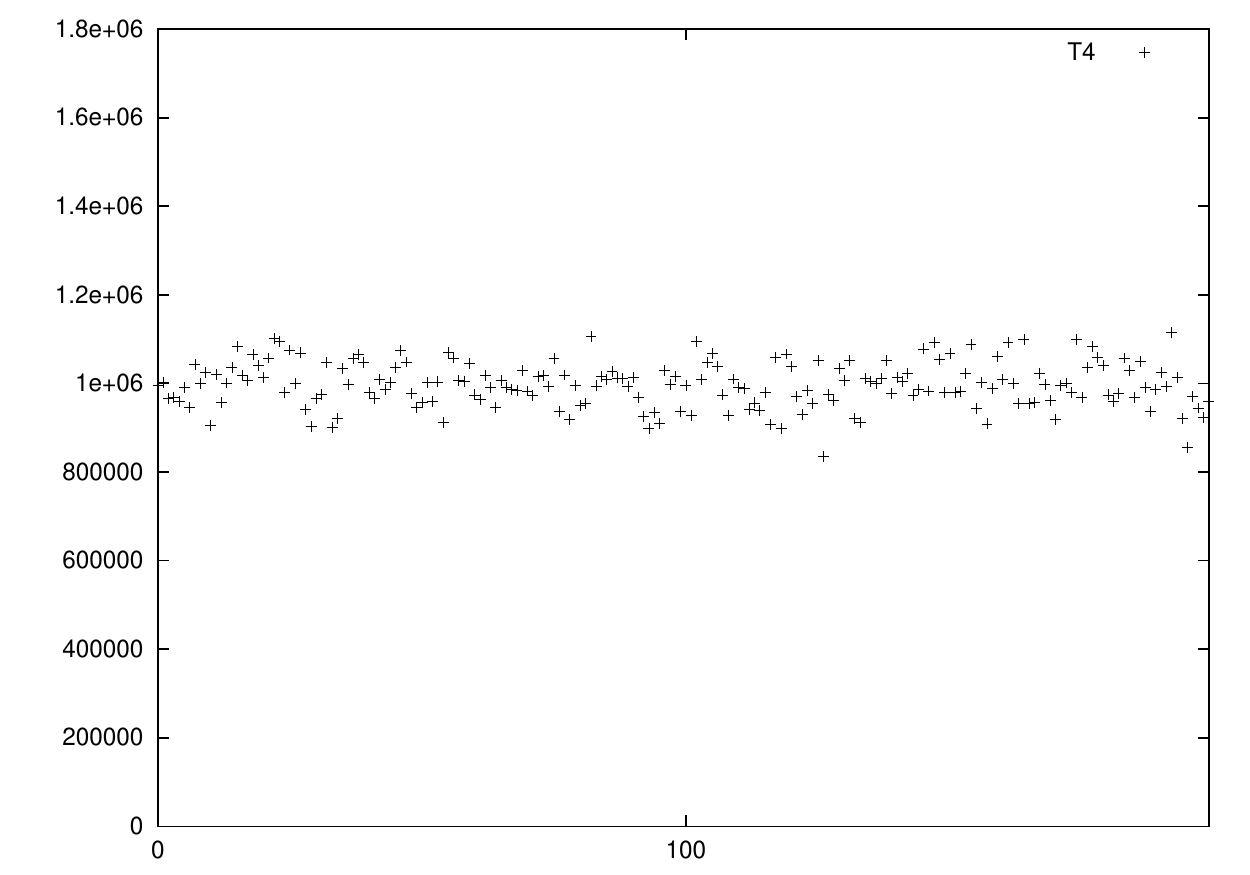}
      \caption{First scenario: we consider four tasks, with a normally distributed number of cycles. After approximately 160 frames, the average task length raises up for three tasks. Vertical axis is the number of cycles, horizontal axis is the frame id, or the time.\label{fig:scen_norm}}
   \end{center}
\end{figure*}

In the following simulation results, we will present two different scenarios. This first one is very simple and smooth, the second one uses data coming from experimental measurement on video decoding platforms. 
For the first scenario, we assume 4 tasks, using a normal distribution for execution length. We consider two phases: the first one, for which characteristics are known by the system, and which contains roughly 160 frames, and the second phase (40 frames), in which three out of the four tasks change their behaviour, and increase their average and worst execution time. This 200-frames scenario is of course run several hundreds of times to obtain usable statistics. Figure \ref{fig:scen_norm} shows graphically an example of actual job execution number of cycles. 

The reason we use such a short scenario (200 frames) is that the ``critical interval'' where jobs are killed is most of the time very short: if we consider a scenario close to the saturation, only a very few number of jobs need to be killed before the system knows the new worst case execution times, and does not have to kill jobs any more (unless the load is too high and the schedulability cannot be guaranteed anymore). And this number of killed tasks only depends of the way the transition of the two phases happens, and not of the length of those two phases.

\begin{figure*}
   \begin{center}
   \newcommand{\lw}{0.24\linewidth}
      \includegraphics[width=\lw]{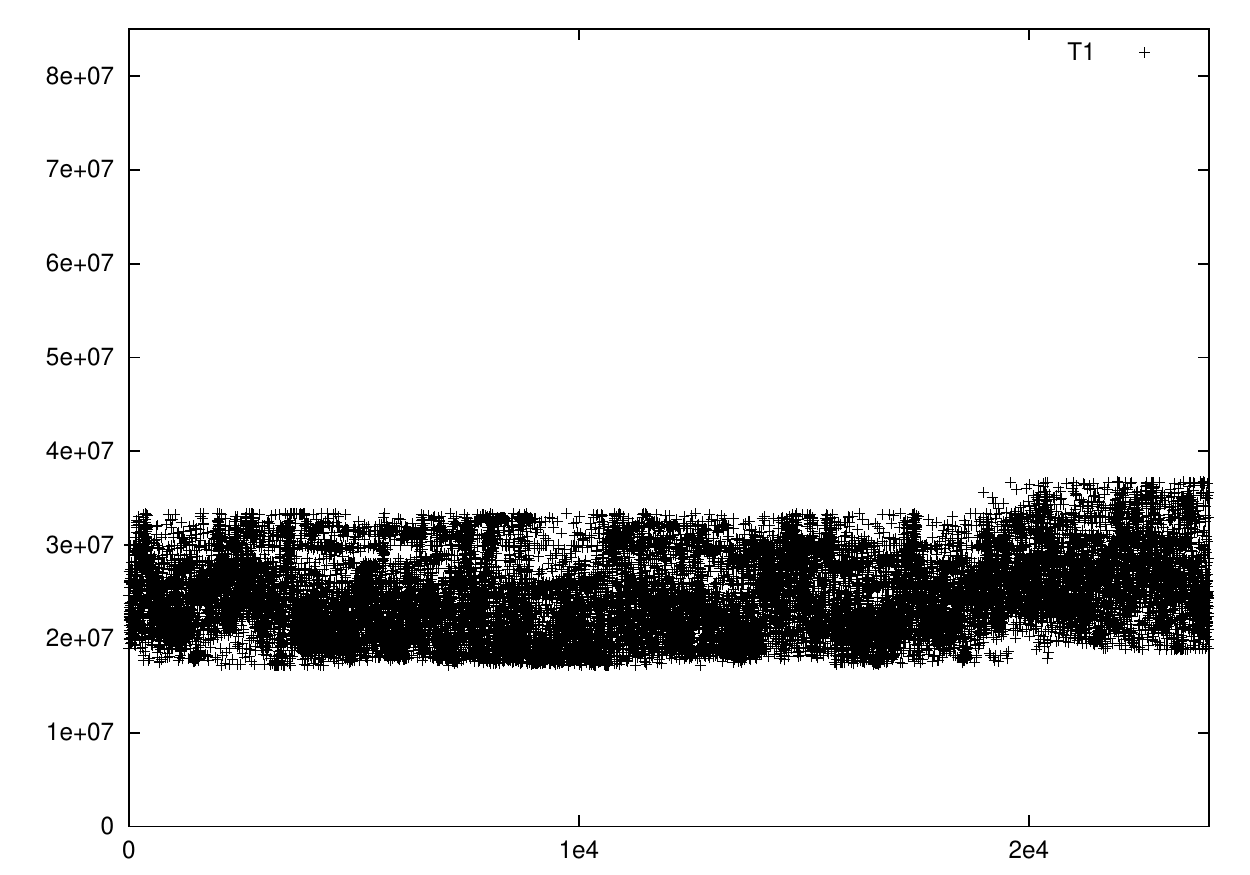}
      \includegraphics[width=\lw]{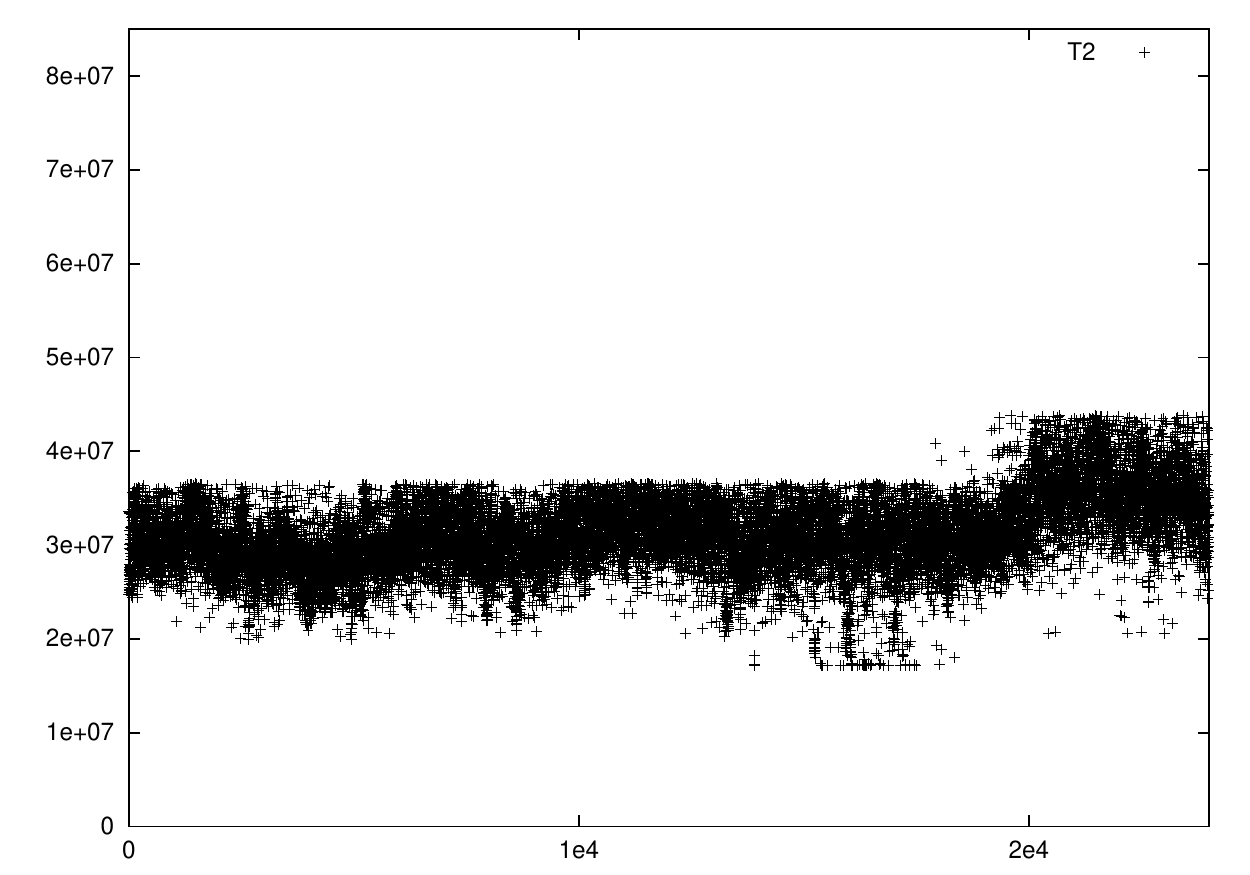}
      \includegraphics[width=\lw]{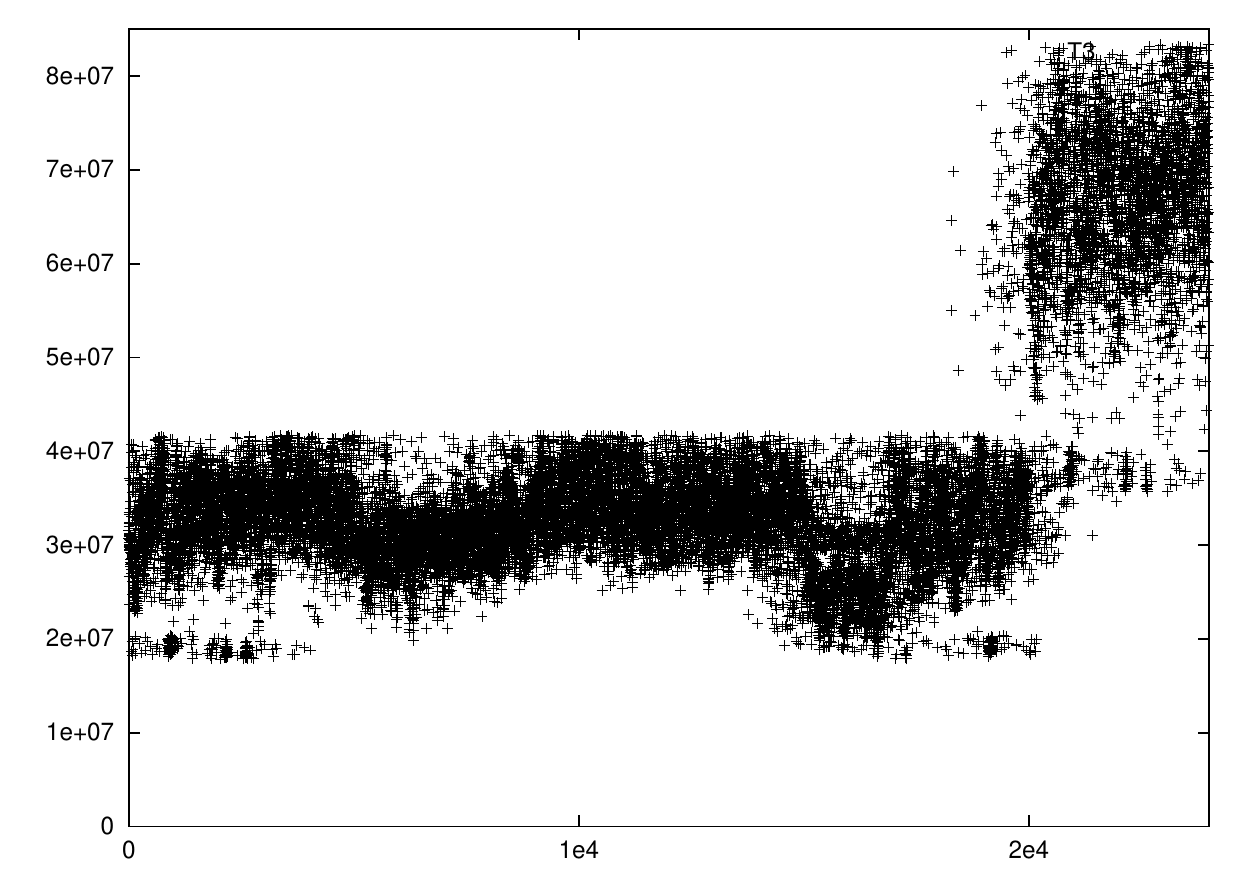} 
       \includegraphics[width=\lw]{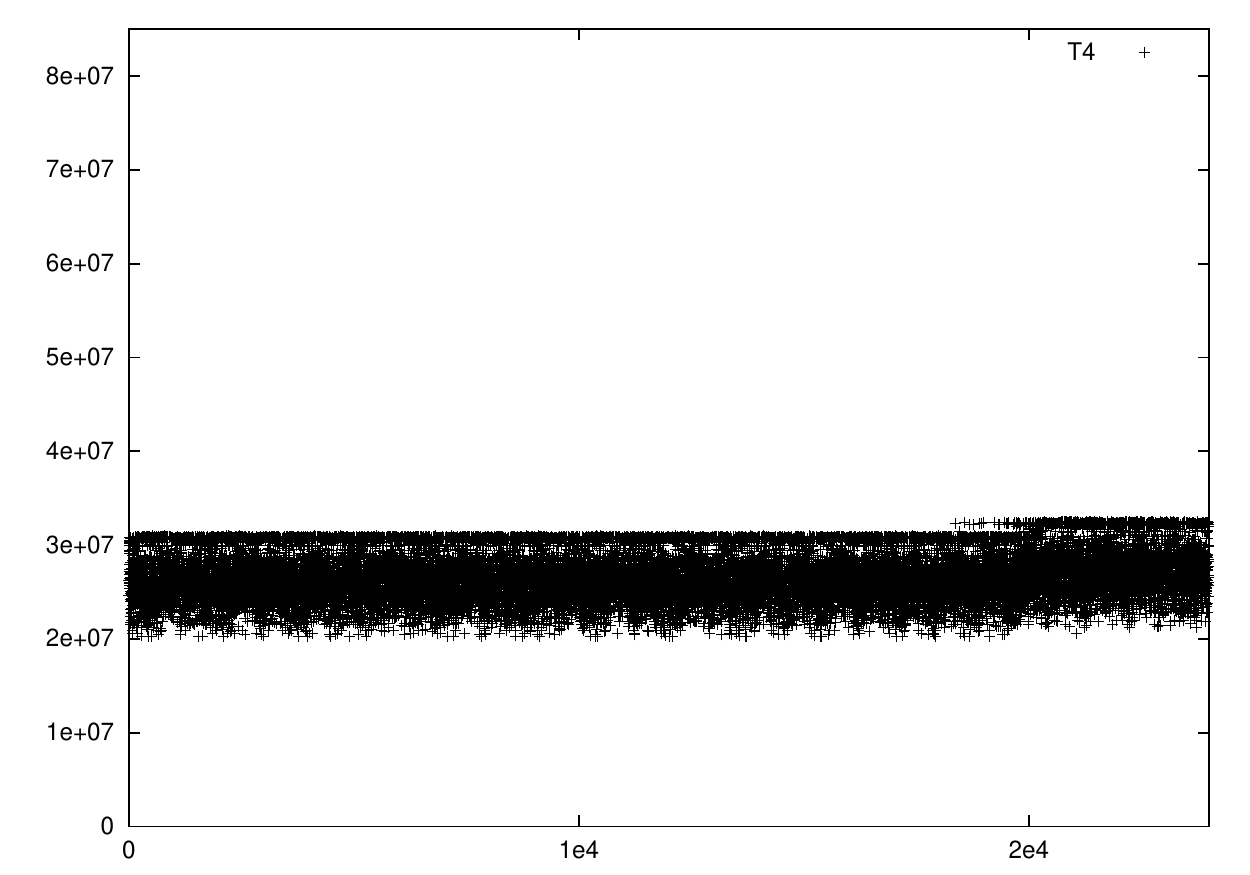}
      \caption{Second scenario (\emph{DIV workload}): we consider nine tasks (we only present four of them here), using a number of cycles observed in an experimental environment (see \cite{RTCSA08}).\label{fig:scen_DIV}}
   \end{center}
\end{figure*}

The second scenario uses data collected on experimental environment in the Computer Science and Information Engineering department of the National Taiwan University, Taipei. We consider 9 tasks, and assume that they are distributed according to the workload presented in \cite{RTCSA08}. Again, we consider two phases, but much longer: the first phase contains 20.000 frames, the second one 4000. See Figure~\ref{fig:scen_DIV}. We refer this scenario as the \emph{DIV workload}.

As the second scenario contains a huge amount of frames, the effect of the transition (number of killed job/lost cycles before the S-functions are up-to-date) is very small. In order to highlight this effect, we also present a short version of this scenario, keeping only 250 frames amongst 20.000.

For every following figure but Fig. \ref{fig:scen_DIV} and \ref{fig:scen_norm}, the horizontal axis represent the frame length. This means that to produce such a plot, we simulated the behaviour of a system with a large frame length (or deadline), measure some metric, and start the same experiment with a shorter frame length. Large value of frame length corresponds then to a low load, and small value to a high load.

\subsection{Fairness}
Measuring the fairness experimentally is not very easy, because for most cases, killing a job should be something quite rare.
We first proposed to measure the laxity of a job as:

$$
\mathcal{L}_i = \frac{1}{\text{nb instances of }T_i} \sum_{\text{instance $k$ of $T_i$}} \frac{e_k}{r_k}
$$
where $e_i$ is the number of cycles that instance $i$ actually run, and $r_i$ is the number of cycle that instance $i$ required. The lower $\mathcal{L}_i$, the higher the number of lost cycles.
We define the fairness as:

$$
\frac{\min_i\{\mathcal{L}_i\}}{\max_i\{\mathcal{L}_i\}}
$$

The drawback of this measurement is that a strategy which kills a very few jobs but always the same (for instance the last one) will have a fairness very close to $1$, while a strategy which kills more often, but different jobs will have higher fairness. However, intuitively, and fair strategy should be a strategy which, when jobs are killed, each task has a similar probability to be the victim. 
So we propose then 

$$
\mathcal{L}_i = \frac{1}{\text{nb killed of }T_i} \sum_{\text{killed instance $k$ of $T_i$}} \frac{e_k}{r_k}
$$
Notice that at low loads, the number of killed jobs is usually very small, then the fairness is computed on a small number of jobs, which gives pretty much erratic values.

\subsection{No preemption mechanism}

\begin{figure}[!ht]
\begin{center}
\includegraphics[width=\linewidth]{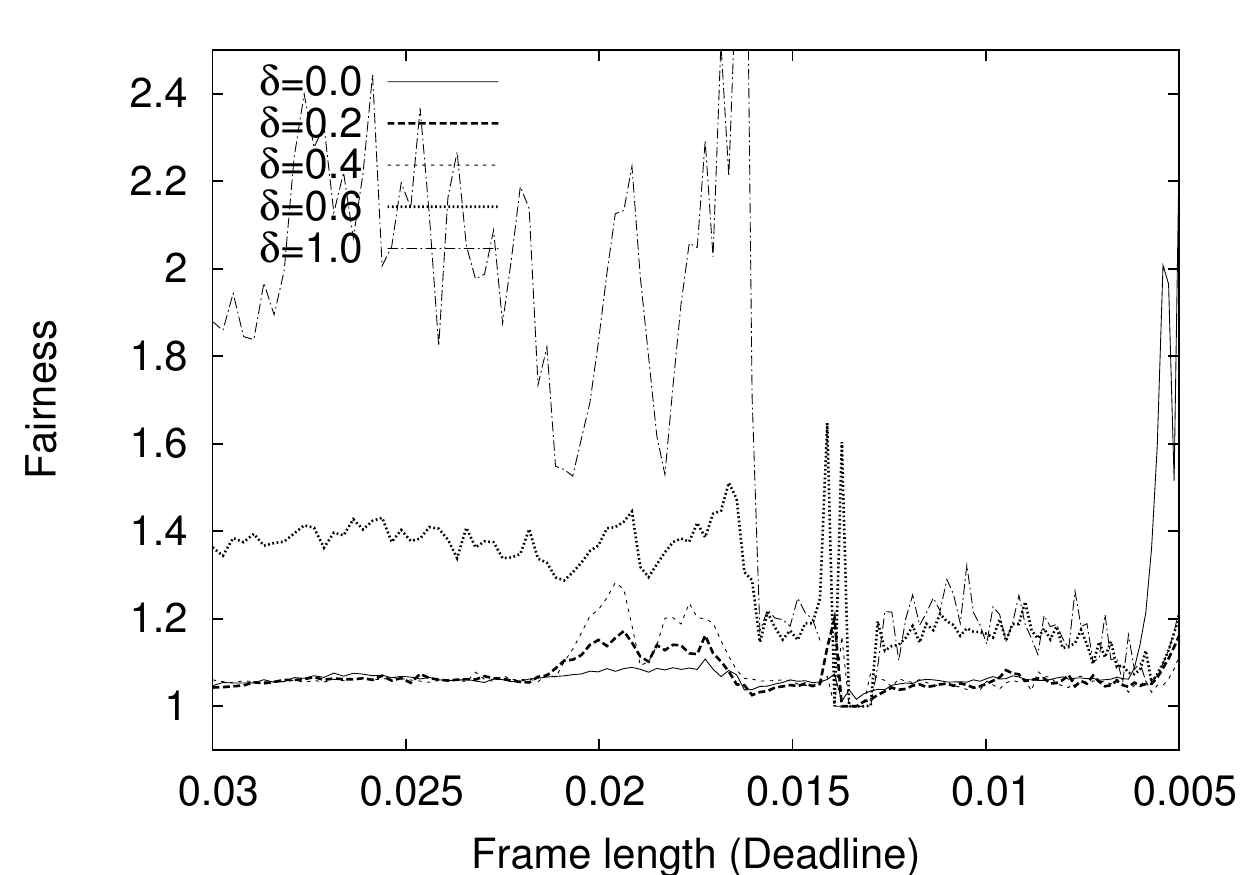}
\caption{Fairness, without preemption mechanism, with four tasks (normally distributed length). The closer to 1, the fairer.\label{fig:norm_nopreempt_fairness}}
\end{center}
\end{figure}

\begin{figure}[!ht]
\begin{center}
\includegraphics[width=\linewidth]{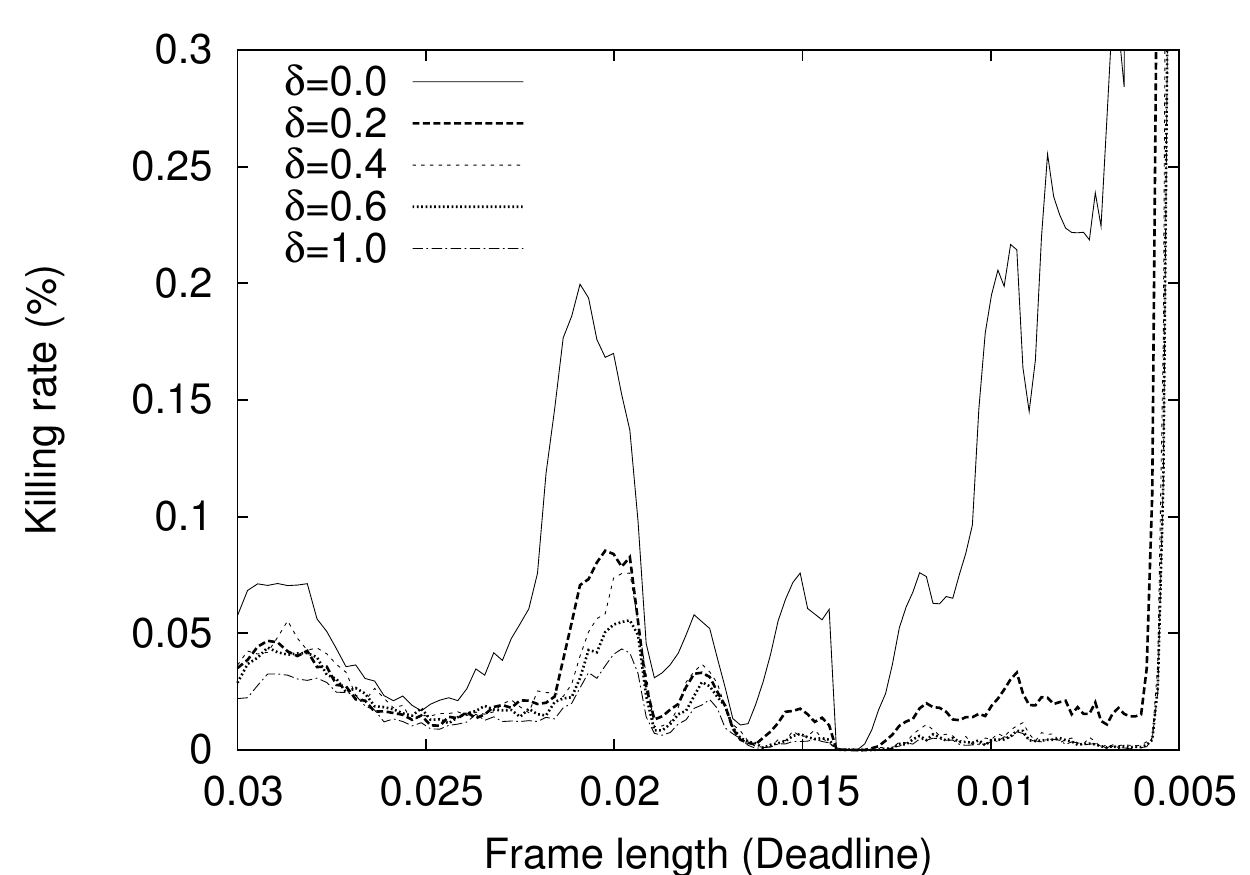}
\caption{Killing rate, without preemption mechanism, with four tasks (normally distributed length)\label{fig:norm_nopreempt_kill}}
\end{center}
\end{figure}

\begin{figure}[!ht]
\begin{center}
\includegraphics[width=\linewidth]{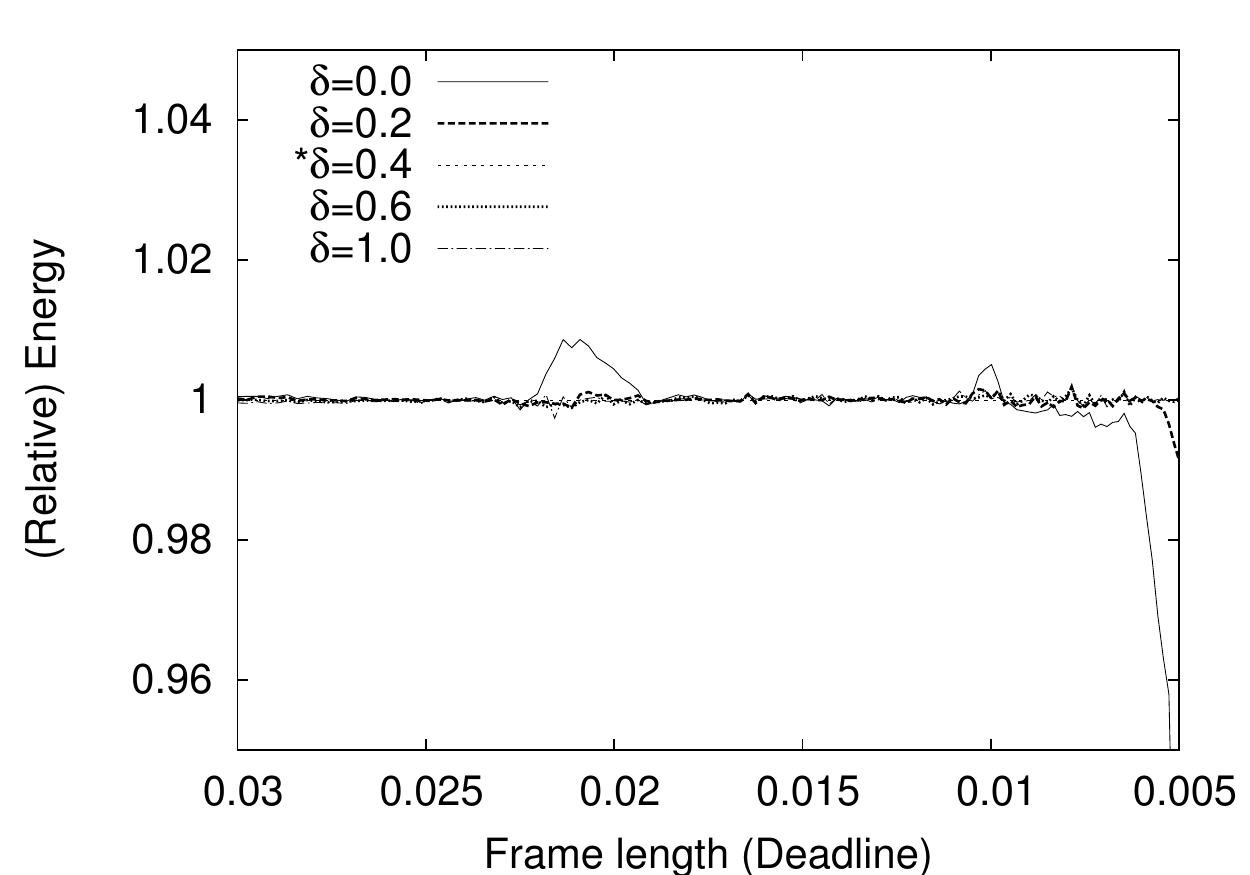}
\caption{Energy consumptions, without preemption mechanism, with four tasks, and normally distributed length\label{fig:norm_nopreempt_energy}}
\end{center}
\end{figure}

In this first set of simulations, using the workload corresponding to Figure \ref{fig:scen_norm}, we assume that we do not have any preemption mechanism, and that we are then only allowed to kill a job and not to resume it. We show in Figures \ref{fig:norm_nopreempt_fairness}, \ref{fig:norm_nopreempt_kill}, and \ref{fig:norm_nopreempt_energy} the influence of the factor $\delta$ --- which gives the laxity we authorise between the danger zone of the next frame, and the end of the frame --- on the fairness, the killing rate and the energy consumption.

It is not surprising that if no flexibility is given (jobs are killed at the next danger zone, $\delta=0$), the fairness (Fig.~\ref{fig:norm_nopreempt_fairness}) is much better (values close to 1) than if we give a large flexibility ($\delta=1$), because in this later case, last jobs are more likely to be killed than first jobs. On the other hand, being more ``rigid'' increases the number of killed jobs at the end of any frame, because we waste some free time that would be available at the end.
So choosing an appropriate value of delta should be done carefully, according to the kind of workload we consider, but also depending whether we have to give more importance to fairness of killing rate. In the example we give here, $\delta=0.2$ seems to be a good trade-off: the fairness (Fig.~\ref{fig:norm_nopreempt_fairness}) if very close to $\delta=0$, but we are comparable to $\delta=1$ regarding to the killing rate (Fig.~\ref{fig:norm_nopreempt_kill}). From the energy point of view (Fig.~\ref{fig:norm_nopreempt_energy}), we don't see any significant difference between different $\delta$'s.

Remark that on the right side of the energy plot (Fig.~\ref{fig:norm_nopreempt_energy}), we see a very huge difference between $\delta$'s. But at this load, a huge number of jobs intrinsically need to be killed, because the frame length is no small to allow tasks to finish. So in most case, designer are not really interested in the system behaviour at those loads.

\subsection{Preemption mechanism}

\begin{figure}[!ht]
\begin{center}
\includegraphics[width=\linewidth]{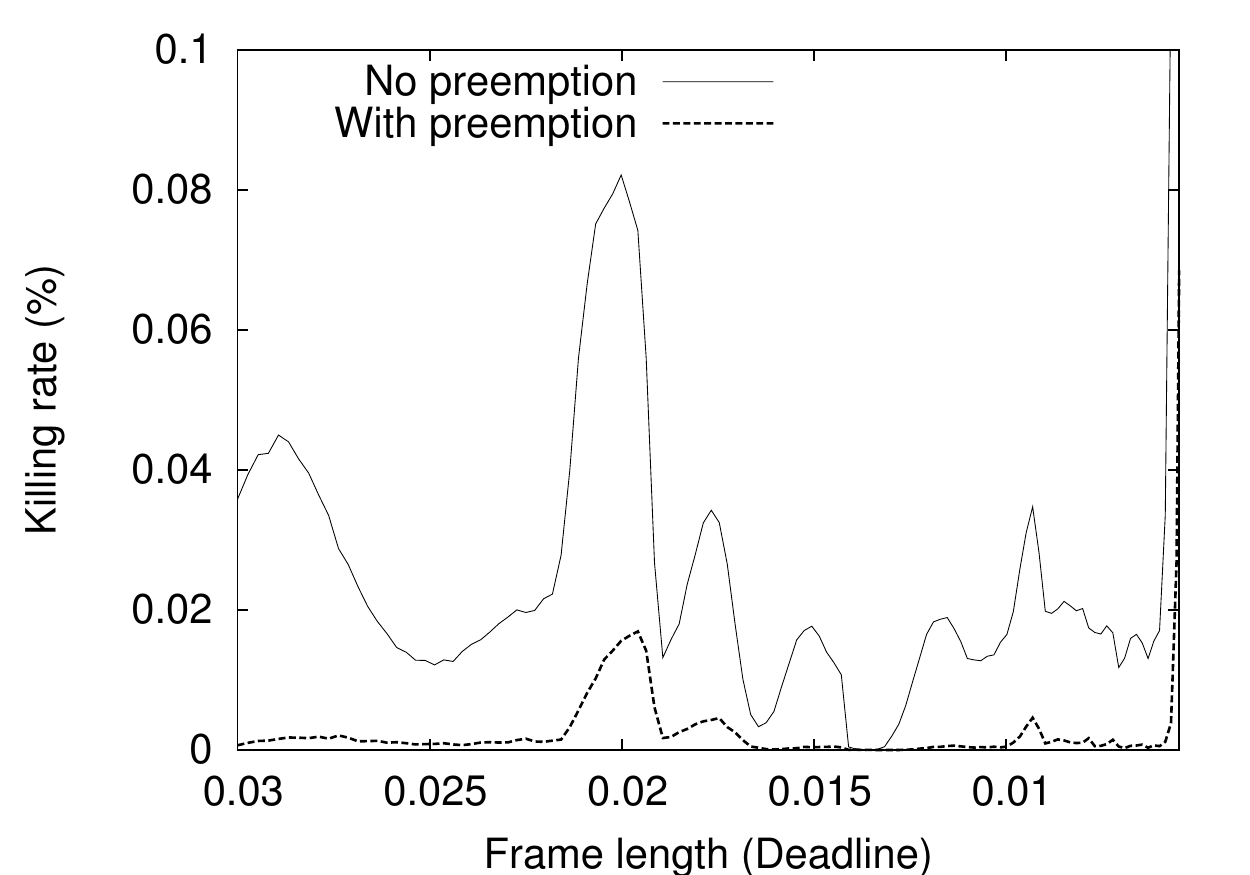}
\caption{Comparison between using or not using preemption mechanism, with four tasks (normally distributed length), and $\delta=0.2$.\label{prempt-nopreempt}}
\end{center}
\end{figure}

First, we compare the difference between having or not a preemption mechanism (Figure \ref{prempt-nopreempt}). Pretty much obviously, we observe that the killing rate is lowered when preemption/resuming mechanism is available. We show for instance a simulation with the normal distribution and $\delta=0.2$. Notice that if we consider high $\delta$, we don't see any difference between using or not a preemption mechanism, because if $\delta$ is high enough, jobs are almost never suspended, and only the last one in the frame is killed.

\begin{figure}[!ht]
\begin{center}
\includegraphics[width=\linewidth]{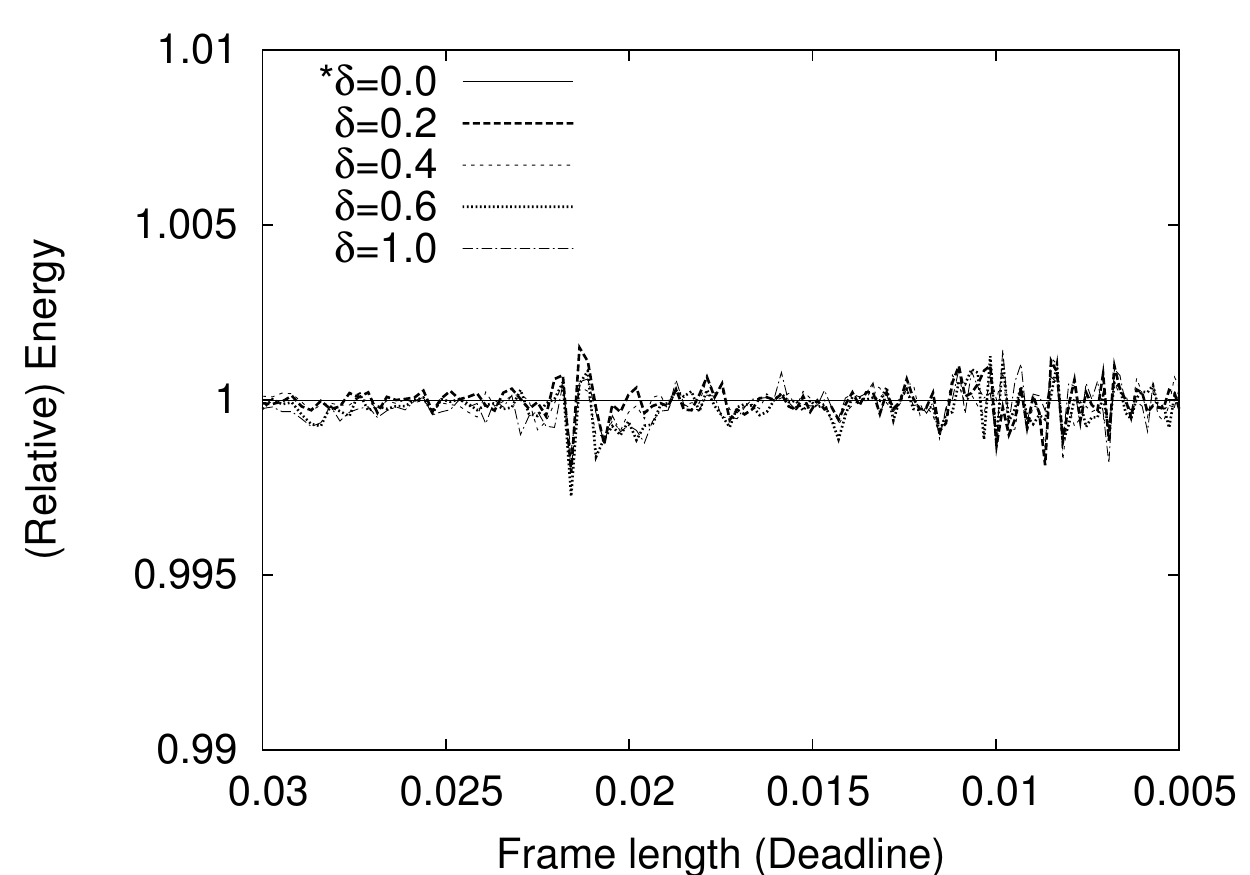}
\caption{Relative energy consumption, with respect to $\delta=0$, with four tasks (normally distributed length), with preemption mechanism (resume at slack)\label{prempt-energy-S}}
\end{center}
\end{figure}

\begin{figure}[!ht]
\begin{center}
\includegraphics[width=\linewidth]{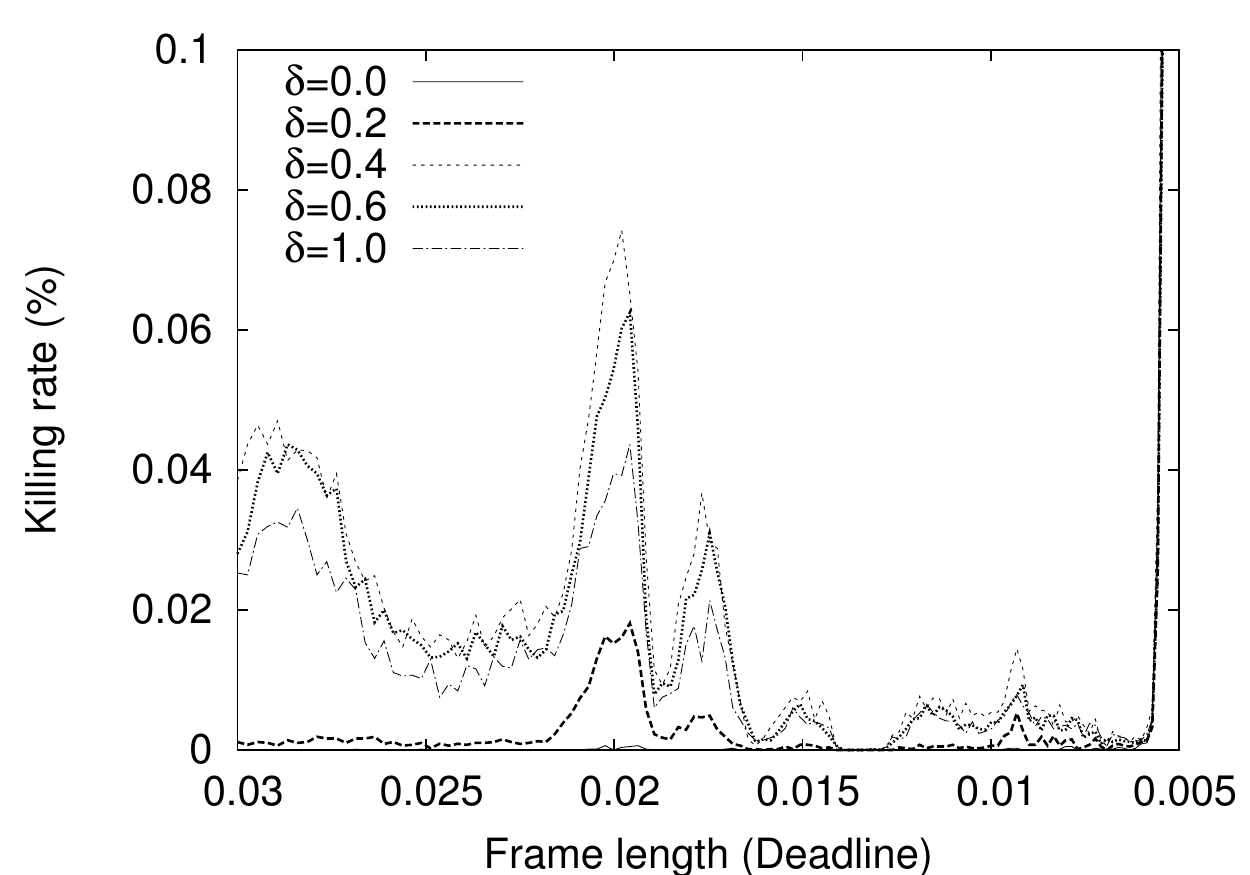}
\caption{Killing rate, with four tasks (normally distributed length), and preemption mechanism (resume at slack)\label{prempt-lost-S}}
\end{center}
\end{figure}

\begin{figure}[!ht]
\begin{center}
\includegraphics[width=\linewidth]{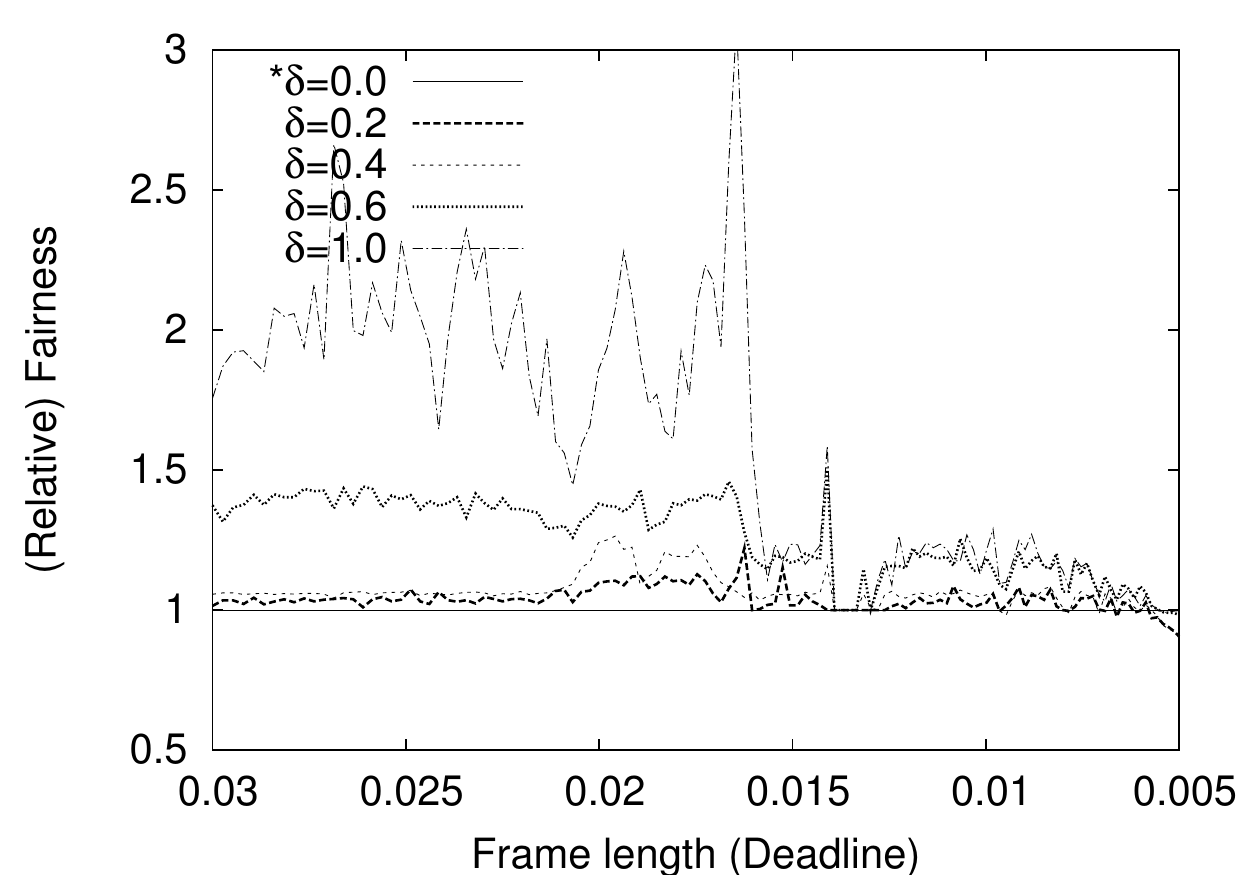}
\caption{Fairness, with four tasks (normally distributed length), and preemption mechanism (resume at slack)\label{prempt-fairness-S}}
\end{center}
\end{figure}

In Figures \ref{prempt-energy-S}, \ref{prempt-lost-S} and \ref{prempt-fairness-S}, we show respectively the energy consumption, average number of cycles, and fairness for various $\delta$ factors, when preemption mechanism is available, and using the normal distribution workloads. We can see in this set of simulation that there is no need to let jobs running in the danger zone: we are better to suspend them as soon as they enter the next danger zone ($\delta=0$), and resume them when some slack time become available.


\subsection{Realistic workload}

In the few next plots, we will show some results where a more realistic workload (see Figure \ref{fig:scen_DIV}) has been used for simulations. We aim here at comparing our adaptation method to two other scenarios: first, a simple case where the scheduling function is not adapted when the workload changes, then an idealistic situation where we know in advance when the change occurs, and we also know the distribution of the two phases.

\begin{figure}[!ht]
\begin{center}
\includegraphics[width=\linewidth]{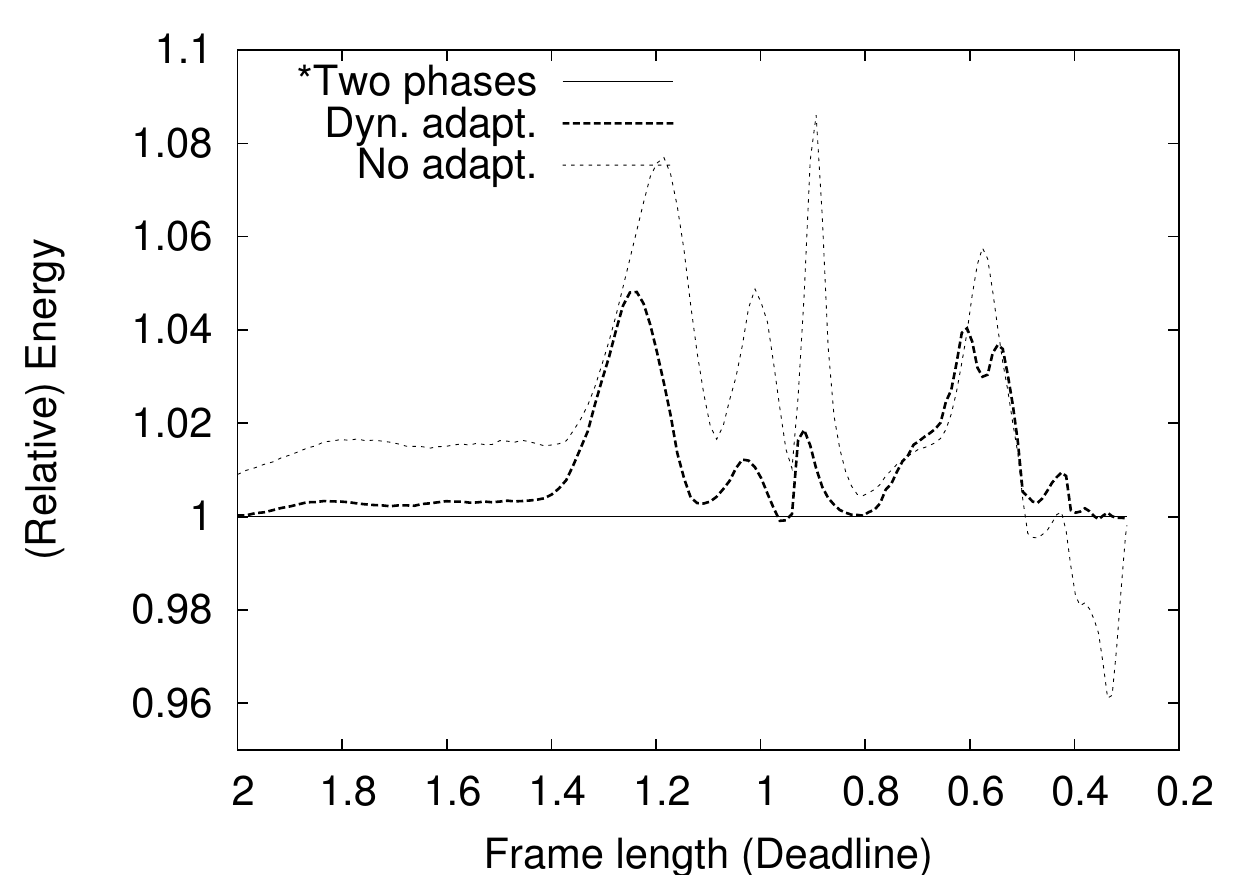}
\caption{Energy consumption, with 9 tasks (DIV workload). We compare our dynamic adaptation mechanism with a clairvoyant method (Two phases), and a simple method which does not adapt its information when the length raises up.\label{DIV-energy}}
\end{center}
\end{figure}

In the first figure (Figure~\ref{DIV-energy}), we observe that our adaptation method saves much more energy that no adaptation at all. The very low energy consumption we can see for the no adaptation scenario at high load comes from the high rate of killed jobs for this method. Indeed, all of those lost cycles do not consume energy at all.

\begin{figure}[!ht]
\begin{center}
\includegraphics[width=\linewidth]{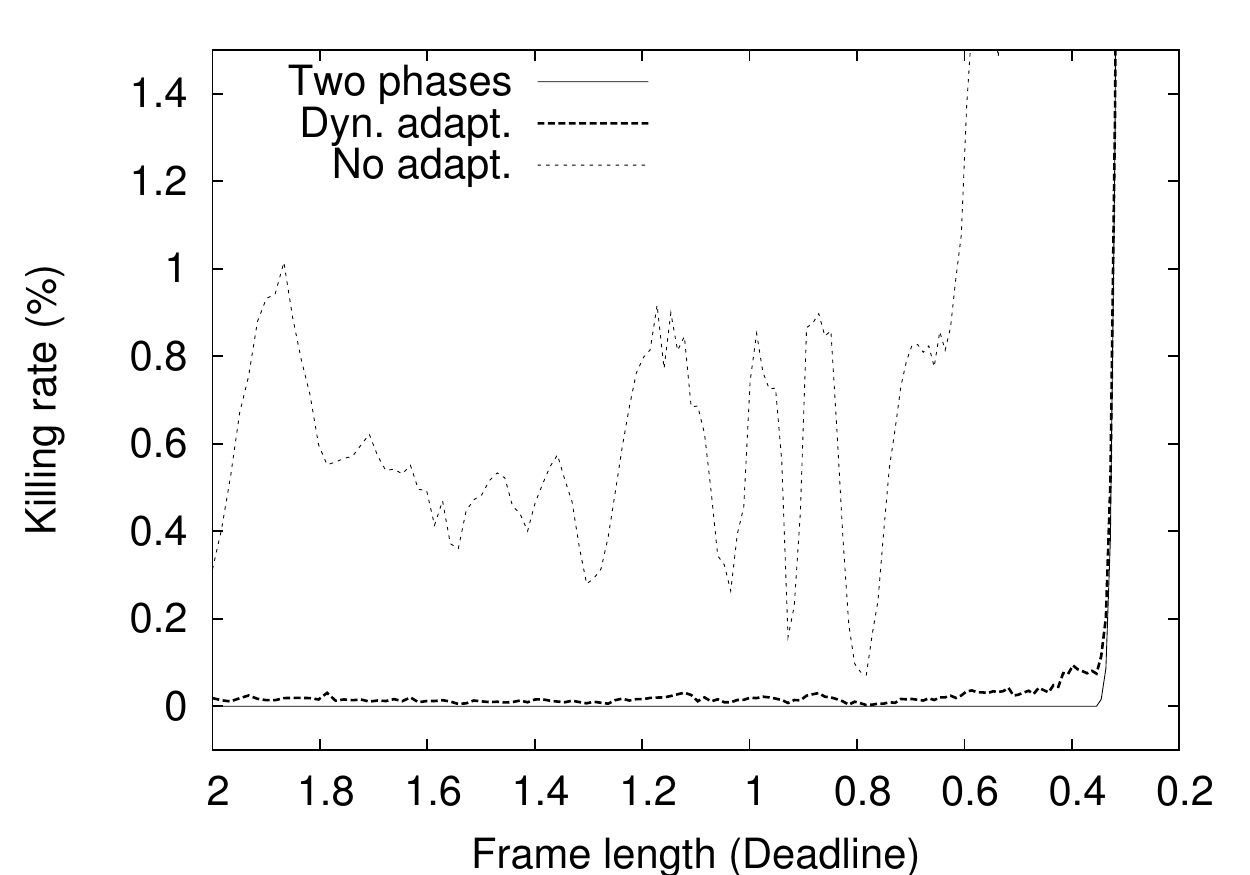}
\caption{Killing rate, with 9 tasks (DIV workload). Same comparison as for Figure\ref{DIV-energy}.\label{DIV-killrate}}
\end{center}
\end{figure}

In Figure \ref{DIV-killrate}, we show the average killing rate for those three scenarios. However, if we consider the same workload as before, we could not see any difference between our dynamic method and the ``2 phases'' method, because our adaptation only needs to kill a few jobs, and this number of jobs does not depends upon the length of the simulation. It's why in this plot, we present measurements done on a much shorter workload ($\pm$ 250 frames). Please notice that the killing rate in the ``no adaptation'' scenario was not affected significantly by the length of the simulation, as long as we keep the same ratio between the first and the second phase.
We observe in this simulation that the killing rate in our adaptation can be very close to zero, which means that even if we do not have a good knowledge of the distribution (and its worst case execution time), we can still, with very small effort, avoid to kill most jobs that would have been killed if we needed to collect a new distribution before adapting the $S$-functions.

\subsection{Miscellaneous experiments}
We have also performed many experiments we do not present in details here, by lack of space. Here are a few conclusions we have drawn:
\begin{itemize}
   \item We almost have not observed any difference between ``resume at end'' and ``resume at slack''. Slack can be used very often, but when slack used, we already entered danger zone in most cases, and then often the frequency is $f_{max}$ up to the end.
   \item The two adaptation methods we proposed (using schedulability condition or horizontal shift) do not show significant difference, whatever the metric we consider. 
   \item Idem for the adaptations we proposed for the danger zone: they both show the same performance or fairness.
\end{itemize}

\section{Conclusion}

In this paper, we have shown that with a small effort, we can efficiently manage varying WCEC on DVS frame-based systems. We provide several algorithms and methods allowing to first have an efficient behaviour as soon as some task overpasses its WCEC, and secondly, adapt the scheduling functions to improve the schedulability. We provide several proofs showing the correctness of our algorithms, and present many simulation results attesting the performance of the proposed methods.
Through those simulations, we shown that we can be very close to a clairvoyant algorithm, both from the killing rate point of view, and from the energy consumption point of view.

\bibliographystyle{acm}
\nocite{*}

\bibliography{VaryingWCEC}
\end{document}